\documentclass[useAMS,usenatbib]{mn2e}

\usepackage[T1]{fontenc}
\usepackage{aecompl}

\usepackage{graphicx}
\usepackage{amsmath}
\usepackage{color}
\usepackage[final]{hyperref}
\usepackage{amssymb}
\usepackage{booktabs}
\usepackage{xspace}

\newcommand{\myemail}{skielboe@dark-cosmology.dk}

\newcommand{\nar}{New Astronomy Reviews}
\newcommand{\apj}{ApJ}
\newcommand{\apjl}{ApJ}
\newcommand{\apjs}{ApJ}
\newcommand{\mnras}{MNRAS}
\newcommand{\pasp}{PASP}
\newcommand{\araa}{ARA\&A}
\newcommand{\aj}{AJ}

\newcommand{\aplett}{Astrophysical Letters}
\newcommand{\nat}{Nature}

\newcommand{\trli}{\ifmmode \tau_{\rm RLI \xspace} \else $\tau_{\rm RLI}$\fi \xspace}
\newcommand{\tccf}{\ifmmode \tau_{\rm CCF \xspace} \else $\tau_{\rm CCF}$\fi \xspace}
\newcommand{\tjav}{\ifmmode \tau_{\rm JAVELIN \xspace} \else $\tau_{\rm JAVELIN}$\fi \xspace}

\newcommand{\javelin}{{\sc javelin \xspace}}
\newcommand{\dnest}{{\sc dnest3 \xspace}}

\newcommand{\halpha}{\ifmmode {\rm H}\,{\alpha \xspace} \else H\,{$\alpha$}\fi \xspace}
\newcommand{\hbeta}{\ifmmode {\rm H}\,{\beta \xspace} \else H\,{$\beta$}\fi \xspace}
\newcommand{\hgamma}{\ifmmode {\rm H}\,{\gamma \xspace} \else H\,{$\gamma$}\fi \xspace}
\newcommand{\civ}{\ifmmode {\rm C}\,{\sc iv \xspace} \else C\,{\sc iv}\fi \xspace}
\newcommand{\mgii}{\ifmmode {\rm Mg}\,{\sc ii \xspace} \else Mg\,{\sc ii}\fi \xspace}
\newcommand{\nii}{\ifmmode [{\rm N}\,{\sc ii}],~{\lambda6586}\,{\rm \AA} \else [N\,{\sc ii}],~$\lambda6586$\,\AA\fi \xspace}

\newcommand{\kms}{\ifmmode {\rm km}\,{\rm s}^{-1} \xspace \else km\,s$^{-1}$\fi \xspace}

\begin{document}

\title[Constraints on the BLR from regularized inversion]{Constraints on the broad line region from regularized linear inversion: Velocity--delay maps for five nearby active galactic nuclei}

\author[A. Skielboe et al.]
{Andreas Skielboe$^1$\thanks{E-mail: \myemail},
Anna Pancoast$^2$,
Tommaso Treu$^{2,3}$\thanks{Packard fellow},
Daeseong Park$^4$\thanks{EACOA fellow},
\newauthor
Aaron J. Barth$^5$,
and
Misty C. Bentz$^6$ \\
$^1$Dark Cosmology Centre, Niels Bohr Institute, University of Copenhagen, Juliane Maries Vej 30, 2100 Copenhagen, Denmark \\
$^2$Department of Physics, University of California, Santa Barbara, CA 93106, USA \\
$^3$Division of Astronomy and Astrophysics, University of California, Los Angeles, CA 90095, USA \\
$^4$National Astronomical Observatories, Chinese Academy of Sciences, Beijing 100012, China \\
$^5$Department of Physics \& Astronomy, 4129 Frederick Reines Hall, University of California, Irvine, CA 92697-4575, USA \\
$^6$Department of Physics and Astronomy, Georgia State University, Atlanta, GA 30303, USA
}

\maketitle

\begin{abstract}
Reverberation mapping probes the structure of the broad emission-line region (BLR) in active galactic nuclei (AGN). The kinematics of the BLR gas can be used to measure the mass of the central supermassive black hole. The main uncertainty affecting black hole mass determinations is the structure of the BLR. We present a new method for reverberation mapping based on regularized linear inversion (RLI) that includes modelling of the AGN continuum light curves. This enables fast calculation of velocity-resolved response maps to constrain BLR structure. RLI allows for negative response, such as when some areas of the BLR respond in inverse proportion to a change in ionizing continuum luminosity. We present time delays, integrated response functions, and velocity--delay maps for the \hbeta broad emission line in five nearby AGN, as well as for \halpha and \hgamma in Arp 151, using data from the Lick AGN Monitoring Project 2008. We find indications of prompt response in three of the objects (Arp 151, NGC 5548 and SBS 1116+583A) with additional prompt response in the red wing of \hbeta. In SBS 1116+583A we find evidence for a multimodal broad prompt response followed by a second narrow response at 10 days. We find no clear indications of negative response. The results are complementary to, and consistent with, other methods such as cross correlation, maximum entropy and dynamical modelling. Regularized linear inversion with continuum light curve modelling provides a fast, complementary method for velocity-resolved reverberation mapping and is suitable for use on large datasets.
\end{abstract}

\begin{keywords}
    galaxies: active -- galaxies: nuclei -- methods: statistical
\end{keywords}

% ============================================================
% ============================================================
% ============================================================
% INTRODUCTION
\section{Introduction} \label{sec:introduction}
% ============================================================

Active galactic nuclei (AGN) are believed to be powered by the release of gravitational potential energy when matter falls onto supermassive black holes in the centres of galaxies. Some AGN have broad emission lines that are thought to be Doppler broadened emission from gas orbiting the central black hole. The broad lines vary in response to the continuum emission, suggesting that they are powered by ionizing radiation originating in the immediate vicinity of the black hole \citep{1969ApJ...155.1077B,1972ApJ...171..213D}. The time delay between the variations in the continuum and emission lines can be used to measure the structure and characteristic size of the broad emission line region (BLR) by the method of reverberation mapping \citep{1972ApJ...171..467B,1982ApJ...255..419B,1993PASP..105..247P}.

The radius of the BLR combined with the width of the emission line provides a measurement of the mass of the central black hole \citep{1998ApJ...501...82P,2000ApJ...533..631K}. Reverberation masses have been found to correlate with properties of the host galaxy, such as stellar velocity dispersion \citep[e.g.][]{2010ApJ...716..269W} and bulge mass \citep[e.g.][]{2009ApJ...694L.166B,2013ARA&A..51..511K}, as has been found for black holes in inactive galaxies \citep[e.g.][]{1995ARA&A..33..581K,1998AJ....115.2285M,2000ApJ...539L...9F,2000ApJ...539L..13G} and indicating a connection between supermassive black hole growth and galaxy evolution. In addition, BLR radii have been found to correlate with the UV/optical luminosity of the AGN \citep[e.g.][]{,2000ApJ...533..631K,2005ApJ...629...61K,2009ApJ...697..160B,2013ApJ...767..149B}, suggesting that AGN can be used as independent standard candles for cosmology \citep{2011ApJ...740L..49W,2014MNRAS.441.3454K}. The radius-luminosity relation also enables black hole masses to be estimated out to large redshift ($z>6$) by measuring the AGN luminosity and broad emission line width in a single spectrum, yielding single-epoch black hole mass estimates \citep[e.g.][]{2002ApJ...571..733V,2002MNRAS.337..109M,2006ApJ...641..689V,2013BASI...41...61S}.

Because of limited knowledge of the structure of the BLR, it is necessary to introduce a scaling factor in traditional reverberation mapping black hole mass measurements, such that the black hole mass is determined by
\begin{equation} \label{eq:ffactor}
    M_\textrm{BH} = f \frac{\Delta V^2 R_\textrm{BLR}}{G},
\end{equation}
where $\Delta V$ is the velocity width of the varying part of the emission line, $G$ is the gravitational constant, and $f$ is the order-unity scaling factor that encapsulates the unknown details regarding the BLR gas and kinematics \citep{1999ApJ...526..579W,2004ApJ...615..645O}. The $f$-factor is generally calibrated using the $M_\textrm{BH} - \sigma_\star$ relation between the black hole mass and the stellar velocity dispersion of the bulge in the host galaxy. This relies on determining $M_\textrm{BH}$ independently in quiescent galaxies using stellar and gas dynamics (see \citealt{2013ARA&A..51..511K} for a review), and then determining the average scaling factor $\langle f \rangle$ that brings the ensemble of active galaxies into agreement with the quiescent $M_\textrm{BH} - \sigma_\star$ relation. This process only corrects for the average offset, but it results in black hole masses that are uncertain for any particular AGN by a factor of $2-3$. In addition, $\langle f \rangle$ is determined using \hbeta, but often applied to other emission lines such as \mgii $\lambda 2799$ and \civ $\lambda 1549$, although the structure of the BLR for these emission lines is very uncertain \citep{2006ApJ...647..901M,2008AJ....135.1849W,2007ApJ...659..997K,2014ApJ...795..164T}.

The main uncertainty in measuring black hole masses using reverberation mapping thus comes from our limited knowledge of the kinematics and geometry of the BLR. A precise measurement of the kinematic and geometric structure of the BLR would enable the determination of $f$ for any individual AGN. This could help reduce scatter in black hole mass determinations for all single epoch mass estimates and their applications \citep[][]{2013ApJ...764...45K}. A better understanding of changes in BLR structure with luminosity and redshift will also reduce systematic errors possibly affecting current scaling relations. To make a full map of the BLR we need to determine not just its characteristic radius, but also the full velocity-resolved transfer function that describes the relation between the continuum emission and the emission line response.

To first order, the problem of reverberation mapping can be formulated as a deconvolution problem in which the flux in the emission-line light curve, at a given wavelength $\lambda$, $F_l(t,\lambda)$ is given by a convolution of the AGN continuum light curve, over some wavelength range, $F_c(t)$ with a transfer function $\Psi(t,\lambda)$ that encodes the physics and geometry of the BLR,
\begin{equation} \label{eq:continous_transferequation}
    F_l(t,\lambda) = \int_{-\infty}^\infty \Psi(\tau,\lambda) F_c(t-\tau)d\tau.
\end{equation}
The transfer function, as a function of time delay and wavelength $\Psi(\tau, \lambda)$, is called the velocity--delay map \citep{1991ApJ...379..586W}. In this approximation, the transfer function is assumed to be linear. Even though the detailed physics is likely to be more complicated, the linear approximation is currently sufficient given that observational datasets have only recently become good enough to probe the full velocity-resolved transfer function. Thus, the transfer function introduced here represents an observed projection of the underlying structure of the AGN. A sound inference of the transfer function will need to account for any residual mismatches between the assumed model and the data, such as non-linearities.

Because transfer functions represent projections of the underlying physical structure, physical and geometrical models are required to interpret them. Several groups have gone through the exercise of predicting transfer functions based on underlying physical models for the BLR structure \citep[e.g.][]{1996ApJ...466..704C,1997ApJ...479..200B,2012MNRAS.426.3086G,1992MNRAS.256..103P}. These studies provide a valuable catalogue of transfer functions that can be consulted when interpreting results from reverberation mapping.

Much effort has gone into developing efficient methods for estimating the BLR size and transfer function $\Psi(\tau,\lambda)$. Early attempts relied on estimating the time delay by-eye \citep{1973ApL....13..165C}, but many sophisticated methods have been developed since. \cite{1982ApJ...255..419B} were the first to calculate transfer functions directly using the convolution theorem of Fourier transforms and provided a comprehensive catalogue of transfer functions for a number of idealised BLR structures. Unfortunately, the requirement for very high quality data, as well as difficulties in dealing with measurements errors, mean that the method has seen little application since its publication. This may change with future high cadence, high signal-to-noise reverberation mapping campaigns.

Another type of inversion procedure is the maximum entropy method that finds the solution for the transfer function that has the highest entropy, while still providing a good fit to the data \citep{1984MNRAS.211..111S,1991ApJ...371..541K,1991ApJ...367L...5H,1994ASPC...69...23H}. Maximum entropy has been successful in estimating transfer functions and velocity--delay maps in a number of AGN \citep[e.g.][]{1991ApJ...371..541K,1996MNRAS.283..748U,2010ApJ...720L..46B,2013ApJ...764...47G}. The downsides of maximum entropy are that it is computationally expensive, it relies on a number of assumptions about the shape of the transfer function, and it is difficult to carry out rigorous error analysis and model comparisons.

Another class of method is dynamical modelling in which a full physical model of the BLR is constructed, and its parameters are inferred within the framework of Bayesian statistics. The statistical framework allows for rigorous error analysis as well model selection \citep{2011ApJ...730..139P}. Furthermore, dynamical modelling circumvents the need for interpreting transfer functions by providing direct estimates of physical model parameters such as inclination and the black hole mass, which in turn allows for the $f$-factor to be calculated directly. The main challenges of dynamical modelling methods are that they require long computation times and the assumption that the model is flexible enough to provide a good description of the BLR.

Dynamical modelling and maximum entropy both provide useful and complementary constraints on the BLR structure. They also rely on a number of assumptions about the allowed shape of the transfer function and are fairly computationally expensive. It is therefore worthwhile to consider alternative methods for analysing reverberation mapping data that are less computationally expensive and allow for greater flexibility in estimating velocity-resolved transfer functions.

Here we develop a method for reverberation mapping based on regularized linear inversion \citep[RLI;][]{1994PASP..106.1091V,1995ApJ...440..166K}, which we extend by including statistical modelling of the AGN continuum light curve light curves. The method is complementary to other reverberation mapping techniques, and has the advantage that it provides an analytical expression for the transfer function with very few assumptions about its shape. RLI is a flexible, free-form method, allowing the data to suggest the form of the inferred transfer function. This makes RLI an ideal tool for exploring BLR physics beyond the framework of current BLR models. At the same time RLI is one of the fastest reverberation methods that provides a direct estimate of the transfer function. We extend the RLI method to include error in the light curves by a combination of Gaussian process modelling and bootstrap resampling, thereby providing a robust estimate of the transfer function and its uncertainties.

As a first application, we apply our method to photometric and spectroscopic light curves of five nearby AGN measured by the Lick AGN Monitoring Project 2008 collaboration \citep[LAMP 2008;][]{2008ApJ...689L..21B}. The main purpose of this project was to measure masses of supermassive black holes in 13 nearby ($z<0.05$) Seyfert 1 galaxies \citep{2009ApJ...705..199B}. Besides improving black hole mass estimates, LAMP 2008 has produced a medley of scientific results including: AGN variability characteristics \citep{2009ApJS..185..156W}, an update of the $M_\textrm{BH} - \sigma_\star$ relation with reverberation mapped AGN \citep{2010ApJ...716..269W}, detailed reverberation mapping studies \citep{2010ApJ...716..993B,2010ApJ...720L..46B}, probing the $R_\textrm{BLR}-L$ relation in the X-rays \citep{2010ApJ...723..409G}, recalibrating single-epoch black hole masses \citep{2012ApJ...747...30P} and dynamical modelling of the \hbeta BLR \citep{2014MNRAS.445.3073P}.

We analyse five objects from LAMP 2008: Arp 151 (Mrk 40), Mrk 1310, NGC 5548, NGC 6814 and SBS 1116+583A, providing integrated response functions, time delays and velocity--delay response maps for the \hbeta emission line in each object, as well as \halpha and \hgamma in Arp 151. We begin in Section \ref{sec:data} by describing how we obtained light curves from the LAMP 2008 dataset. In Section \ref{sec:methods} we outline the RLI method and show results from simulations. Section \ref{sec:results} presents the main results of our analysis. The results are discussed in Section \ref{sec:discussion} and finally we provide a short summary and conclusions in Section \ref{sec:summary-and-conclusion}.

All time delay-axes in the figures are in the observed frame. All time delays quoted in the text and in Table 1 are in the rest frame of the AGN.

% ============================================================
% ============================================================
% ============================================================
% DATA
\section{Data} \label{sec:data}
% ============================================================

We use data from LAMP 2008\footnote{Data available at \url{http://www.physics.uci.edu/~barth/lamp.html}}, a dedicated spectroscopic reverberation mapping campaign that ran for 64 nights at the Lick Observatory 3-m Shane telescope. The spectroscopic data were supplemented by photometric monitoring using smaller telescopes, including the 30-inch Katzman Automatic Imaging Telescope (KAIT), the 2-m Multicolor Active Galactic Nuclei Monitoring telescope, the Palomar 60-inch telescope, and the 32-inch Tenagra II telescope. A detailed description of the LAMP 2008 observing campaign and initial results are published in \cite{2009ApJS..185..156W} and \cite{2009ApJ...705..199B}.

\subsection{Continuum light curves}
For our analysis, we use Johnson $B$ and $V$ broad band continuum light curves from LAMP 2008. The bands were chosen to improve dynamical modelling results and resolve variability features. The fluxes are measured using standard aperture photometry (see \citealt{2009ApJS..185..156W} for a complete discussion). The light curves are the same as those used in dynamical modelling by \cite{2014MNRAS.445.3073P}. The $B$ and $V$ band light curves are very similar for all objects analysed \citep[see][]{2009ApJS..185..156W}, and the choice of continuum light should not significantly affect our results.

\subsection{Emission line spectra and light curves} \label{sec:emission-line-lightcurves}
Emission-line light curves are generated from flux-calibrated spectra from the LAMP 2008 campaign. The \hbeta emission line is isolated in all objects using spectral decomposition, by modelling all line and continuum components individually, and subtracting away everything but the emission line of interest \citep{2012ApJ...747...30P}. We decide to keep the narrow component of \hbeta to avoid introducing additional error at the centre of the line. This should not affect our results, as the narrow-line component is constant on the time-scale of the observing campaign \citep[see][]{2012ApJ...747...30P}, and because we only consider variations around the mean flux in the line. In Arp 151 we also analyse the \halpha and \hgamma lines, where \hgamma is isolated using spectral decomposition in a way similar to that for \hbeta \citep{2011ApJ...743L...4B}. Because no spectral decomposition is available for \halpha we follow the procedure of \cite{2010ApJ...716..993B}, isolating \halpha by subtracting a straight line fit to two wavelength windows on either side of the line.

The resulting emission line spectra have the same spectral resolution as the original LAMP 2008 data. The spectral dispersions are provided in \cite{2009ApJ...705..199B} and range from $11.6 - 14.7$\,\AA\,\,(FWHM), corresponding to $5.9 - 7.5$ pixels per resolution element.

A few of the spectra in the LAMP 2008 dataset have been identified as unreliable due to low spectral quality or suspicious features above the noise \citep{2012ApJ...747...30P}. We tested the effect of removing these spectra in the analysis and found that it made little difference to the overall results. Because of this, and to be able to compare directly with recent results from dynamical modelling \citep{2014MNRAS.445.3073P}, we decided to include the unreliable spectra in the RLI analysis presented here. The emission line light curves for \hbeta are thus identical to those used for the dynamical modelling analysis in \cite{2014MNRAS.445.3073P}.

% ============================================================
% ============================================================
% ============================================================
% METHODS
\section{Regularized linear inversion} \label{sec:methods} \label{sec:rli}
% ============================================================

Regularized linear inversion seeks to solve the transfer equation (Equation \ref{eq:continous_transferequation}) for the transfer function $\Psi(\tau,\lambda)$ analytically and without any assumptions regarding its functional form (only that it is a bounded linear operator). This approach is potentially very powerful in that it relies solely on the data when deriving the transfer function. This in turn allows for very little freedom for the method to select a proper solution. In the presence of noise, or if the system deviates slightly from the linear assumption, this lack of freedom means that a solution cannot be found at all. Therefore, instead of looking for a unique solution that fits the data, the method determines the solution that minimizes the $\chi^2$. A simple $\chi^2$ minimization will generally over-fit the data and make the inversion unstable. This is overcome by a regularization in which the first-order derivative of the solution is to be minimized along with the $\chi^2$, such that the solution is smoothed at the level of the noise \citep{phillips1962technique,tikhonov1963solution}. The result is comparable to maximizing the entropy under a $\chi^2$ constraint as in the case of maximum entropy methods. RLI has several advantages over other methods, specifically it 1) makes no assumption about the shape or positivity of the transfer function, 2) can be solved analytically, and 3) has very few free parameters (the regularization scale as well as the transfer function window and resolution). Note that while RLI can in principle fit any shape of transfer function, in practice the result will be limited by the sampling and noise in the data, which means that a minimum scale exists below which the response function will be unresolved.

By measuring the continuum ($F_c$) and emission line ($F_l$) light curves we can in principle solve the transfer equation \eqref{eq:continous_transferequation} for $\Psi(t,\lambda)$. In reality, data is always discrete, so we rewrite the transfer equation as a linear matrix equation of the form
\begin{equation} \label{eq:discrete_transferequation}
    \mathbf{L}_{\Delta \lambda} = \mathbf{\Psi}_{\Delta \lambda} \mathbf{C},
\end{equation}
where $\mathbf{L}_{\Delta \lambda}$ is the emission line light curve integrated over the wavelength range (spectral bin) $\Delta \lambda$, $\mathbf{C}$ is a matrix of continuum light curves (see below), and $\mathbf{\Psi}_{\Delta \lambda}$ is the transfer function corresponding to the given wavelength range. While we allow the transfer function to depend on wavelength, symmetries in the BLR as well as observational projections will tend to correlate transfer functions in the time and wavelength domains. The set of transfer equations over a range of frequencies across an emission-line we call the velocity--delay map for the given emission line \citep{1991ApJ...379..586W}.

For perfect noise-free data solving the discrete transfer equation \eqref{eq:discrete_transferequation} would be a simple matter of inverting $\mathbf{C}$ to obtain the transfer function $\mathbf{\Psi}$. Because of noise we cannot hope to find an exact solution to the linear inversion problem. Instead we seek a solution that minimizes the $\chi^2$ together with a smoothing condition that ensures that we are not fitting the noise.

Interpreting the transfer function to make statements about the structure of the BLR relies on a number of assumptions. First, we assume that the variations in the AGN continuum bands are correlated with the AGN ionizing continuum (this is not necessarily true, see e.g. \citealt{1998ApJ...500..162C,2014MNRAS.444.1469M,2015ApJ...806..129E} who find evidence for a time delay between the UV and optical continuum in NGC 7469 and NGC 5548). Second, we assume that the continuum emission originates from a region negligible in size compared to the BLR. Third, we assume that the continuum ionizing radiation is emitted isotropically. Last, we assume that the BLR structure is constant and the response linear for the duration of the campaign, such that a single linear transfer function can be calculated from the full light curves.

\subsection{Solving for the response function}
Rather than solving for the transfer function, which includes constant emissivity components of the BLR, we consider only variations about the mean of the light curves. By doing this, we are measuring the response of the line emission to a change in the continuum emission that is ionizing the BLR gas. Hence the quantity we are solving for, $\Psi(\tau,\lambda)$, is the response function \citep{1991ApJ...371..541K,1993MNRAS.263..149G}.

Following the notation of \cite{1995ApJ...440..166K}, and considering only variation about the mean of the light curves, we write $\chi^2$ as
\begin{equation} \label{eq:chi2}
    \chi^2 = \sum_{i=M}^N \frac{1}{\sigma_l^2(t_i)} \left [ \delta F_l(t_i) - \sum_{j=1}^M [F_c(t_i - \tau_j) - \langle F_c \rangle]\Psi(t_j) \right ]^2.
\end{equation}
This expression can be recast to matrix notation,
\begin{equation} \label{eq:chi2matrix}
    \chi^2 = (\mathbf{L} - \mathbf{C}\mathbf{\Psi})^2,
\end{equation}
where the light curves enter as
\begin{align}
    \mathbf{C}_{ij} &= [F_c(t-\tau_j) - \langle F_c \rangle] / \sigma_l(t_i) \label{eq:C}\\
    \mathbf{L} &= \delta F_l (t_i) / \sigma_l(t_i), \label{eq:L}
\end{align}
and the variation about the mean in the emission-line light curve is defined as
\begin{equation}
    \delta F_l (t_i) = F_l(t_i) - \langle F_l \rangle.
\end{equation}
We choose a time delay resolution of $1\,\textrm{day}$, as this corresponds to the highest resolution of the data. By running the analysis with different resolutions we have confirmed that the choice of resolution, within reasonable values, does not change our results. Contrary to \cite{1995ApJ...440..166K}, we calculate a constant mean of the continuum light curve data points such that $\langle F_c \rangle$ is not a function of the time delay $\tau$. This is done in order to allow meaningful comparisons between different continuum realizations (see Section \ref{sec:cont-errors}). Minimizing $\chi^2$ in Equation \eqref{eq:chi2matrix} leads to the linear equation
\begin{equation}
    \mathbf{C^T C \Psi} = \mathbf{C^T L}.
\end{equation}
Although this expression looks simple, it turns out to be ill-conditioned. To remedy this, we put an extra constraint on the problem, namely that the solution should be smooth at the scale of the noise (this effectively avoids fitting the noise). To guarantee smoothness of the solution, we introduce a differencing operator $\mathbf{H}$ acting on $\mathbf{\Psi}$, and require that the first-order difference (the discrete version of the first-order differential) be minimized together with the ordinary $\chi^2$. To control the weights between the $\chi^2$ and the first-order difference, a scaling parameter $\kappa$ is introduced that sets the scale of the regularization. Thus, the expression to minimize becomes
\begin{equation} \label{eq:regularization}
    (\mathbf{C^T C} + \kappa \mathbf{H^T H}) \mathbf{\Psi} = \mathbf{C^T L}.
\end{equation}
This expression is more stable under inversion, while sacrificing detail by emphasizing smoothness of the solution. The question is then how to choose a suitable scale $\kappa$ for the regularization.

The best choice of regularization scale depends on the signal-to-noise in the data, as well as the level of uncorrelated systematic uncertainties that result in deviations from the assumption of linear response. As a starting point for selecting a regularization scale $\kappa$, we follow the recommendation of \cite{press1992numerical} in which $\kappa = \kappa_0$ is chosen to provide equal weights to the two left-hand-side terms in equation \eqref{eq:regularization},
\begin{equation} \label{eq:lambda0}
    \kappa_0 = \frac{\textrm{Tr}(\mathbf{C^T C})}{\textrm{Tr}(\mathbf{H^T H})}.
\end{equation}
\cite{1995ApJ...440..166K} suggest using the largest value of $\kappa$ which gives an acceptable $\chi^2$, while \cite{1994PASP..106.1091V} suggest using a value of $\kappa$ which provides the ``best'' trade-off between resolution and noise. By running tests on simulated data, we find that good results are generally achieved for $\kappa = \kappa_0$. For this reason, and to reduce the number of free parameters in the method, we fix $\kappa = \kappa_0$ for all our analysis (see Section \ref{sec:changing-the-regularization-scale} for a test of the effect of changing the regularization scale).

We calculate all response functions in the time delay interval $0 - 30$\,days. The lower bound is chosen to impose causality in the solution. The upper bound is chosen to be well beyond the time-scale where we expect a significant response based on previous measurements and expectations for Seyfert galaxies \citep{2013ApJ...767..149B}. We tested the effect of changing the time delay windows and found that it had only minor effects on the results, as longs as the window is long enough to include the main response. When presenting the results, we show only the first part of the response function from $0 - 15$\,days, as we found the significant response power to be located at these scales.

\subsection{Continuum light curve errors} \label{sec:cont-errors}

Continuum flux errors are not included explicitly in the RLI formalism presented above. Instead, we include continuum errors and interpolation by modelling the continuum light curves using Gaussian processes \citep{2011ApJ...730..139P}. Studies of AGN variability, using sampling intervals of days, have suggested that AGN continuum variability is well modelled by a damped random walk or Ornstein--Uhlenbeck (O--U) process \citep{2009ApJ...698..895K,2010ApJ...708..927K,2010ApJ...721.1014M,2013ApJ...765..106Z}. Formally this is a CAR(1) or continuous-time first-order autoregressive process, which is a stationary Gaussian process with a power spectral density (PSD) slope of $-2$. The CAR(1) process is often used in reverberation mapping methods for modelling the continuum light curve and enable efficient interpolation and error estimation \citep[e.g.][]{2011ApJ...735...80Z,2011ApJ...730..139P}.

Recent high cadence observations of AGN using the \emph{Kepler} spacecraft \citep{2011ApJ...743L..12M,2012ApJ...749...70C,2014ApJ...795....2E,2015MNRAS.451.4328K}, have found steep PSD slopes ($< -2$) inconsistent with a CAR(1) process. The effect of having a steeper PSD slope is to suppress small scale variability in the light curves. Using a CAR(1) process for modelling the continuum light curves may therefore result in over-fitting spurious features, due to noise at short time scales. This introduces artificial structure at the level of the noise, that can lead to underestimated errors being propagated form the continuum model to other reverberation mapping parameters.

The magnitude of the error introduced by assuming the wrong continuum model depends on the sampling rate and errors in the data, as well as how it is implemented in the reverberation mapping method. Regularized linear inversion, as implemented here, mitigates the effects noise in the light curves by smoothing the solution at the shortest time scales. This results in a solution that is dominated by the longer multi-day variability features. For this reason, the response functions derived should not be affected by the details of the assumed continuum model, as long as multi-day time scale features are accurately reproduced. Even so, the shallower slope of the CAR(1) process, compared to \emph{Kepler} AGN light curves, may still lead to underestimated errors.

\begin{figure}
    \includegraphics[width=1.0\columnwidth]{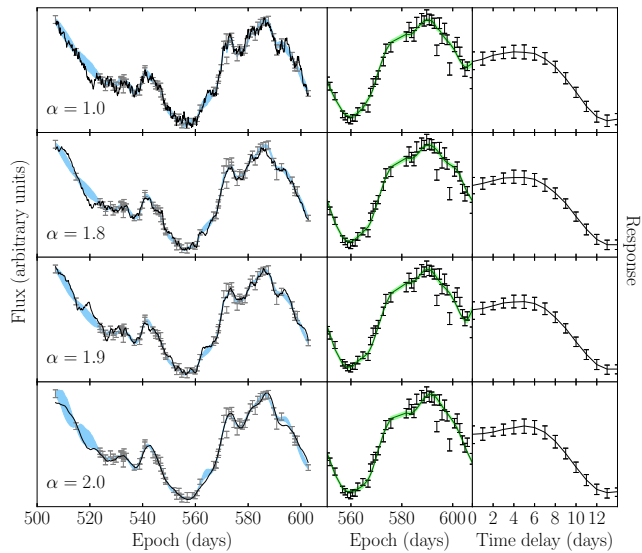}
    \caption{The effect of using continuum models with varying levels of small scale structure. The upper panels show the CAR(1), $\alpha=1.0$ in Equation \eqref{eq:covariance}, continuum model used for the results in this paper. The panels below show results for continuum models with increasing $\alpha$, resulting in decreasing structure at short time scales, corresponding to progressively steeper, or more negative, power spectral density slopes. The left panels show the continuum light curve of Arp 151 from LAMP 2008 as grey error bars. In each case, the best fit (median and 1$\sigma$ percentiles) continuum model is shown as a blue band and a single realization of the best fit continuum model is shown as a solid black line. The middle panels show the Arp 151 \hbeta emission line light curve (black error bars) along with the fit from RLI (green line and band) using the corresponding continuum models on the left. The right panels show the resulting response functions from RLI. The excellent agreement between response functions derived using different continuum models leads us to conclude that the level of short time scale structure in the continuum model does not affect the RLI results derived from the LAMP 2008 dataset.}
    \label{fig:psd-slopes}
\end{figure}

To determine how the choice of continuum model PSD affects our results, we run RLI using a number of different continuum models with varying degrees of small scale structure, corresponding to varying PSD slopes. We do this by changing the power $\alpha$ in the covariance function for the Gaussian process. The covariance is given by \citep{2011ApJ...730..139P}
\begin{equation} \label{eq:covariance}
	C(t_1, t_2) = \sigma^2 \exp \left [ - \left ( \frac{|t_2-t_1|}{\tau} \right )^\alpha \right ],
\end{equation}
where $\sigma$ is the long-term standard deviation, $\tau$ is the typical time scale of variations and the power $\alpha$ take on values in the interval $[1,2]$. A power of $\alpha = 1$ corresponds to the CAR(1) model, whereas $\alpha > 1$ produces smoother continuum models, with PSD slopes $< -2$. Using data from LAMP 2008 for Arp 151 (see results in section \ref{sec:results-arp151}), we test four different continuum models with $\alpha = 1.0$, $1.8$, $1.9$ and $2.0$ respectively (see Figure \ref{fig:psd-slopes}). We find that the emission line fits and response functions are not strongly affected by the assumed continuum model. For this reason, and to be consistent with other reverberation mapping methods such as \javelin and dynamical modelling, we choose to use a CAR(1) process as a prior for the Gaussian process continuum modelling. Future analyses of high-cadence reverberation mapping data should likewise be careful in considering the effects of the assumed continuum model.

The finite sampling and duration of the light curves hampers our ability to model structure on scales close to or below the sampling time scale, and on scales longer than the duration of the light curve. At short time scales the light curves will likely be dominated by observational errors due to the small fractional variability at these scales. Regularized linear inversion deals with this by smoothing the solution at the smallest time scales, resulting in a solution dominated by the longer multi-day variability features. For all the objects analysed, the time scale of interest (a few days) is well within the sampling rate of the data.

For each continuum light curve we find the best fit Gaussian process parameters using the \dnest Nested Sampling code by \cite{2010ascl.soft10029B}. The best-fit Gaussian process is used to interpolate the continuum light curve, which we need to calculate the response for arbitrary time delays.

In addition to interpolation, we use the statistical variability of the Gaussian process to generate a number of realizations of the continuum light curve that are used to estimate statistical errors on the calculated response functions. This allows us to include measurement errors in the continuum light curves that otherwise do not explicitly enter in the RLI formalism. For each continuum light curve, we generate 1000 realizations of the best-fit Gaussian process. This library of continuum realizations is used throughout the analysis to calculate response functions and response maps.

\subsection{Emission-line light curve errors} \label{sec:line-errors}
Emission-line light curve errors $\sigma_l(t)$ are included explicitly in the RLI formalism in eqs. \eqref{eq:C} and \eqref{eq:L}. We allow for the possibility of additional systematic errors in the emission line fluxes by bootstrap resampling of the emission line light curves in each inversion. Bootstrap resampling is done by re-sampling data points in the light curve, such that each point can be chosen zero or more times and the total number of data points is kept constant. If a point is chosen $N$ times, its error is reduced accordingly by $\sqrt{N}$. We find that the results are only weakly affected by bootstrap resampling, indicating that systematic uncertainties affecting individual epochs in our data sample do not have strong influence on the results.

\subsection{Testing on simulated data (1D)}
\begin{figure}
    \includegraphics[width=1.0\columnwidth]{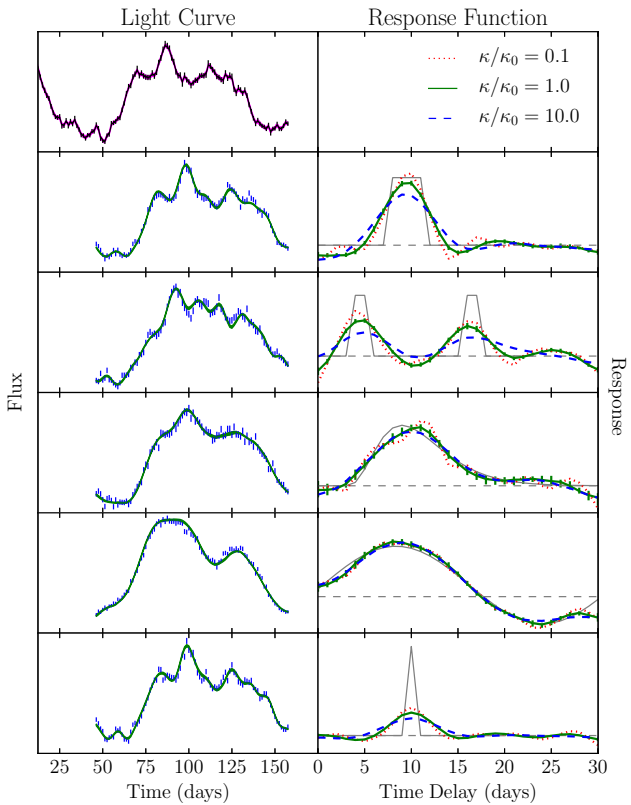}
    \caption{Testing regularized linear inversion on simulated data by recovering response functions for simulated light curves. The top left panel shows the simulated continuum light curve, generated using a Gaussian process with a power spectral density similar to that found by recent Kepler observations \protect\citep{2014ApJ...795....2E}. The magenta band plotted on top of the continuum shows the standard deviation of the Gaussian process fit used in the analysis. The left panels, below the continuum, show the emission line light curves obtained by convolving the continuum with the corresponding response functions shown in the right panels. Before being analysed, the simulated light curves are degraded and down sampled to a level similar to the light curves of LAMP 2008. The response functions recovered from regularized linear inversion are plotted on top of the input response functions in the right panels. The thin black line is the input response function while the solid green (red dotted and blue dashed) lines shows the response functions calculated from regularized linear inversion with a regularization scale of $\kappa/\kappa_0 = 1$ ($\kappa/\kappa_0 = 0.1$ and $\kappa/\kappa_0 = 10$). The response functions are, from top to bottom: top-hat, multimodal top-hat, Gamma distribution, sinusoidal, and the delta function. The grey dashed horizontal lines indicate the level of zero response.}
    \label{fig:sim-1d}
\end{figure}

We test our RLI method on simulated velocity-integrated (1D) light curves convolved with a selection of different response functions to produce a continuum-emission line light curve pair. To make the light curves as realistic as possible, the continuum light curve is simulated using a Gaussian process with a power spectral density similar to that found by recent Kepler observations \citep{2014ApJ...795....2E}. The normalization of the simulated continuum light curve is chosen to match that of typical LAMP 2008 light curves \citep{2009ApJ...705..199B}, such that the simulated continuum has a fractional variability of $F_{\textrm{var}} \sim 12$ per cent \citep[$F_\textrm{var}$;][]{2003MNRAS.345.1271V} and a signal-to-noise ratio of $\textrm{SNR} \sim 100$. Emission line light curves are generated by convolving the simulated continuum light curve with the chosen response function (see Fig. \ref{fig:sim-1d}). The simulated light curves are degraded and down-sampled to a level similar to the LAMP 2008 dataset (see section \ref{sec:data}). We test our code on five different response functions: top-hat, multimodal top-hat, Gamma distribution, sinusoidal, and a delta function. The simulated light curves and response functions are shown together with the fits from RLI in Fig. \ref{fig:sim-1d}. We further test our method by simulating continuum light curves using a damped random walk. Regularized inversion performs slightly better in this case, especially for small values of $\kappa$, due to the increased structure at short time scales, but the overall results are very similar to the simulated Kepler light curves shown in Fig. \ref{fig:sim-1d}.

RLI is generally successful in recovering all simulated response functions, but like all reverberation mapping methods the overall performance depends on the number of strong variability features in the light curve sample. Because of the smoothing constraint and the finite sampling of the light curves, RLI cannot fit very sharp features in the simulated response functions. This is particularly evident in the case of the top-hat and delta function response functions. For the smooth Gamma and sinusoidal distributions, we find that RLI is able to match the full shape of the response function. We tested the method on different simulations and found that the deviations at particular time delays are driven by noise in the input data and thus cannot be reduced without re-sampling, or extending, the light curve.

The smallest scale resolvable by RLI can be seen in the recovery of the delta-function in the bottom panel in Fig. \ref{fig:sim-1d}. This smallest scale, or point spread distribution, comes from the incomplete sampling of the light curves as well as the smoothing imposed on the solution to avoid fitting noise in the input.

We test the effect of changing the regularization scale $\kappa$ by one order of magnitude in each direction. The effect of increasing the regularization ($\kappa/\kappa_0 = 10$, dashed blue line in Fig. \ref{fig:sim-1d}) is to smooth the response functions, thereby reducing some of the fluctuations in the wings, but also significantly increasing the width of the PSF (Fig. \ref{fig:sim-1d}, lower right panel). In the same way, a reduction in regularization scale ($\kappa/\kappa_0 = 0.1$, dotted red line in Fig. \ref{fig:sim-1d}) tends to amplify spurious features in the response functions while improving the PSF by decreasing its width. The best trade-off between high resolution and signal-driven results is a matter of preference. We find that $\kappa = \kappa_0$ provides a good balance between resolution and noise. For this reason, and for consistency, we fix the regularization scale to $\kappa = \kappa_0$ for all our analysis.

It is important to note that these simulations assume an exact linear relation between the continuum and emission line, Equation \eqref{eq:continous_transferequation}, including only random errors and sampling gaps. The simulations and the results from RLI presented in Fig. \ref{fig:sim-1d} thus represent an artificial best-case scenario. The main purpose of Fig. \ref{fig:sim-1d} is to illustrate the effect of changing the regularization scale $\kappa$, show the finite point spread distribution of the RLI method due to discrete sampling, and the ability of RLI to recover negative and multimodal response functions.

\subsection{Testing on simulated data (2D)}
\begin{figure}
    \includegraphics[width=\linewidth]{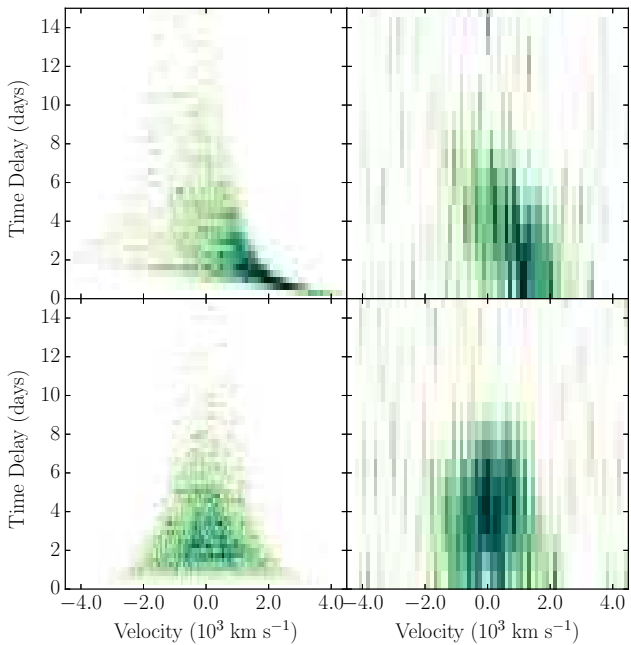}
    \caption{Testing regularized linear inversion on simulated two-dimensional response maps (left column) generated using the dynamical modelling code of \protect\cite{2014MNRAS.445.3055P}. The response functions are calculated for each velocity bin individually using RLI, and then plotted together to produce velocity--delay maps (right column).}
    \label{fig:sim-2d}
\end{figure}

In addition to testing one-dimensional response functions, we test our RLI implementation on data simulated using the geometric and dynamical model of the BLR used in dynamical modelling of reverberation mapping data \citep{2011ApJ...730..139P,2011ApJ...733L..33B}. The dynamical modelling method simulates the broad line region as a number of discrete point particles in a parametrized geometrical configuration. We test RLI on two different simulated BLR configurations, one in which the BLR particles are in near-circular orbits, and one with entirely inflowing orbits (see \citealt{2014MNRAS.445.3055P} for details on the simulated datasets). The results are shown next to the true response maps in Fig. \ref{fig:sim-2d}. We note that the data are simulated using a kinematic and geometric model for the BLR, and are not necessarily representative of true AGN response maps.

These simulations provide a qualitative indication of the 2D point-spread-function of the RLI method. Because the solution is forced to be smooth in the presence of noise, the resulting response maps will not reproduce the input maps exactly. Instead, we retrieve a smoothed version of the input response maps. The stripes in the RLI maps appear because we analyse each velocity bin individually. This means that the solution is not smoothed in the velocity (or wavelength) direction. We did this to simplify the calculations and to let any correlations between velocity bins be driven by the data only. This is contrary to maximum entropy, where the result is also smoothed (or regularized) in the velocity direction. If some velocity bins have exceptionally large errors, RLI might fail in that bin and produce a flat response.

Although the response maps from dynamical modelling and RLI look somewhat different, they share some of the same symmetries. The effect of in-fall in the upper panel in Fig. \ref{fig:sim-2d} is clearly recovered in the RLI map. Likewise, the lower panel in Fig. \ref{fig:sim-2d} shows that dynamical modelling and RLI both show a symmetric response on either side of the emission-line centre, suggesting a structure where the BLR clouds are on closed orbits around the central black hole.

\subsection{Effect of changing the regularization scale \texorpdfstring{$\kappa$}{k}} \label{sec:changing-the-regularization-scale}

\begin{figure}
    \includegraphics[width=\linewidth]{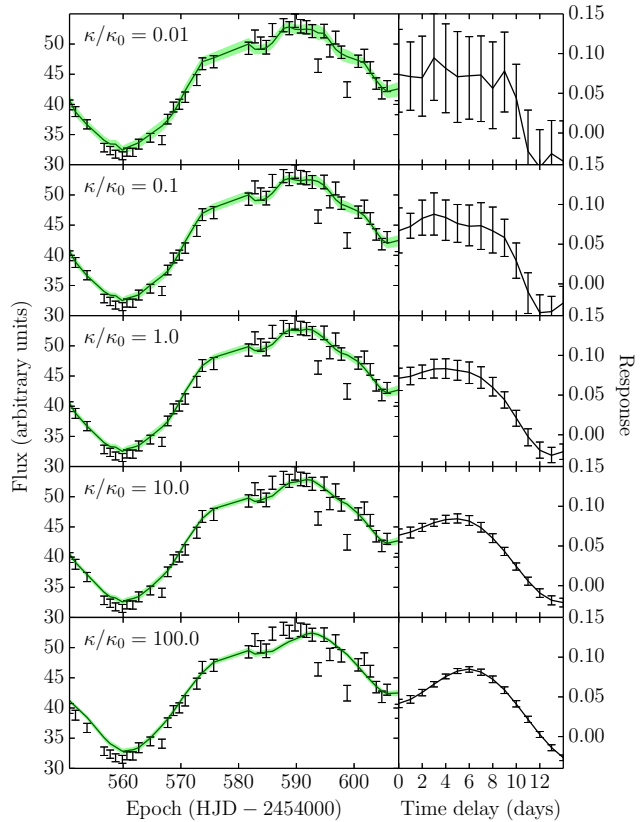}
    \caption{Effect of changing the regularization scale $\kappa$, increasing the smoothness of the solution. Using Arp 151, \hbeta as an example. The left panels show the \hbeta light curve (black points with error bars) along with the fit from RLI (solid green line and band). The right panels show the corresponding response function found using RLI. The regularization scales affects all time scales in the response function equally, which is why the error bars are more of less constant as a function of time delay for each regularization scale $\kappa$.}
    \label{fig:result-changing-lambda}
\end{figure}

Fig. \ref{fig:result-changing-lambda} shows the effect of changing the regularization scale $\kappa$ using the \hbeta data for Arp 151 as an example (see full results in Section \ref{sec:results-arp151}). We find that, for low vales of $\kappa$ (i.e. more weight on the $\chi^2$-term), the solution is very unstable to perturbations in the input, and the error in the response function increases at all time delays. The reason that the error is not a function of time delay, is that the noise is not correlated with the signal in the light curve. Therefore, the response function can draw power at any time delay to fit the noise, so long as it draws similarly less power at other scales. Because the noise is random in each continuum realization, the time delays at which the response function draws its power will be evenly distributed in delay space, and the response function will fluctuate with more or less the same amplitude at all time delays. There will still on average be more power at scales correlated with the signal, thus the median shape of the response function peaks around the true delay, but the amplitude variations due to noise produce large error bars at all time delays. This behaviour is expected and illustrates the need for regularization to achieve stable solutions.

As the regularization scale is increased, more weight is put on the smoothness of the solution. This suppresses the effects of noise in the input and emphasizes the large scale behaviour of the light curves. Because of the smoothing introduced in the regularization, there will be a minimum resolvable scale, analogous to an extended point-spread-function (see Fig. \ref{fig:sim-1d}). Any structure below the noise level will thus be smoothed out in RLI. 

% ============================================================
% ============================================================
% ============================================================
% RESULTS
\section{Results} \label{sec:results}
% ============================================================

Here we present results from regularized linear inversion of light curves of five local AGN from the LAMP 2008 dataset. Fig. \ref{fig:result-arp151} to \ref{fig:result-sbs1116} show results from regularized linear inversion of broad band photometric continuum light curves and \hbeta light curves from spectral decomposition. We provide fits to the light curves as well as integrated response functions and velocity--delay maps for all object. Table \ref{table} lists derived median time delays for the integrated light curves of all AGN analysed, as well as time delays from cross-correlation and \javelin for comparison.

\subsection{Analysis}

\subsubsection{Light curves}

The upper left panels in Fig. \ref{fig:result-arp151} to \ref{fig:result-sbs1116} show the observed light curves plotted as black points with error-bars. The blue band on top of the continuum light curve shows the median and 1$\sigma$ percentiles of 1000 realizations of the Gaussian process best-fit model. This band indicates the range of continuum light curve realizations used together with bootstrap resampling to determine errors on the derived response functions (see Sections \ref{sec:cont-errors} and \ref{sec:line-errors}). The middle left panels show the integrated emission-line light-curves obtained by integrating the corresponding emission-line over the wavelength range listed in Table \ref{table}. The solid line and green band on top of the emission line light curves show the median and 1$\sigma$ percentiles of the fits obtained from RLI when analysing all 1000 realizations of the continuum together with 100 bootstrap resamplings for each continuum realization.

\subsubsection{Integrated response functions}

The lower left panels of Fig. \ref{fig:result-arp151} to \ref{fig:result-sbs1116} show the best-fit integrated response functions under the smoothing constraint described in Section \ref{sec:rli}. The response function and the associated error-bars are calculated from the median and 1$\sigma$ percentiles of RLI using 1000 iterations of continuum light curve interpolations as well as 100 bootstrap resamplings of the emission line light curve. Because we minimize not just the $\chi^2$, but also the first derivative of the response function, the medians and error-bars of the response functions will be naturally correlated.

\subsubsection{Velocity-resolved response functions}

The upper right panels of Fig. \ref{fig:result-arp151} to \ref{fig:result-sbs1116} show the velocity--delay maps from regularized linear inversion of the spectra for all epochs in the data. Each spectral bin is treated individually, such that any correlations below the pixel resolution (see Section \ref{sec:emission-line-lightcurves}) are driven by the data. As in the case of the integrated analysis, we fix the regularization scale to $\kappa = \kappa_0$ for all velocity bins.

The velocity--delay maps show the response on a linear scale as a function of time delay and Doppler velocity with respect to the emission-line centre. White colour corresponds to zero response while black is maximum response (see the colour bars for each individual object). Below the response maps we plot the spectrum of the corresponding emission line. The black line is the mean spectrum across all epochs considered and the grey line is the root-mean-square variability (standard deviation about the mean) of the spectra, normalized to the same scale as the mean spectrum.

The lower right panels of Fig. \ref{fig:result-arp151} to \ref{fig:result-sbs1116} show a selection of response functions for three separate velocity-bins. The location of the bins are indicated by the vertical dashed lines in the response maps and spectra. The colours correspond to the relative wavelengths, such that the red response function is for the highest radial velocity bin (redshifted with respect to the line centre), the green response function is for the line centre at $0~\rm{km}/\rm{s}$ and the blue response function is for the lowest radial velocity bin (blueshifted with respect to the line centre).

\subsubsection{The time delay} \label{sec:the-time-delay}

The scalar time delay from cross correlation methods, \tccf, is typically calculated as the centroid of the CCF above a threshold (usually $80$ per cent of the maximum of the CCF). This effectively ignores any contributions from negative response. To compare our results to those from cross correlation, we calculate a RLI time delay, \trli, as the median of the positive values of the response function. The time delays and error bars for each emission line and object, quoted in Table \ref{table} and in the text, are then estimated as the median and 1$\sigma$ percentiles of all time delays determined from the response functions of all continuum realizations (see Sections \ref{sec:cont-errors} and \ref{sec:line-errors}). All time delays quoted in the text and in Table \ref{table} are in the rest frame of the AGN. The time delay units on the axes of the figures are in the observed frame.

\subsubsection{Comparing with other methods}

We compare our results to those of other methods, including cross-correlation, \javelin, maximum entropy, and dynamical modelling.

% Cross-correlation
The cross-correlation method calculates the cross-correlation function (CCF) between the continuum and emission line light curve to find the time delay, generally characterized by the centroid of the CCF, of the responding gas in the BLR. First applied by \cite{1986ApJ...305..175G}, the method has since been substantially developed \citep[e.g.][]{1988ApJ...333..646E,1994PASP..106..879W,1998PASP..110..660P}. The CCF is related to the transfer function through the light curve auto-correlation function \citep{1991vagn.conf..343P}. This means that the width of the CCF depends on the variability characteristics of the AGN as well as the sampling of the data. For this reason, the CCF is rarely interpreted by itself, but is rather used as a tool to estimate the characteristic size of the BLR by calculating the time delay. Table \ref{table} compares integrated time delays from RLI to time delays determined using cross correlation methods by \cite{2009ApJ...705..199B} for \hbeta and \cite{2010ApJ...716..993B} for \halpha and \hgamma. Fig. \ref{fig:result-misty-bins} compares velocity-resolved time delays from RLI to velocity-resolved time delays from cross correlation of the \hbeta line by \cite{2009ApJ...705..199B}. The comparisons for each object are described in Section \ref{sec:results-for-each-agn}. See the previous section for a description of how the RLI time delays are estimated from the response functions.

% JAVELIN
\javelin finds the time delay by using a top hat model for the transfer function \citep{2011ApJ...735...80Z}. Light curve variability is modelled using a CAR(1) process, which is the same process used to model the continuum variability in our RLI method. Like the cross-correlation method, \javelin provides only a scalar time delay as a measure of the BLR structure. \cite{2013ApJ...773...90G} determined time delays using \javelin on the LAMP 2008 dataset. We list time delays from \cite{2013ApJ...773...90G} in Table \ref{table} alongside results from cross-correlation and RLI for comparison. All \javelin time delays $\tjav$ quoted in the text are also from \cite{2013ApJ...773...90G}.

% Dynamical modelling
The dynamical modelling method \citep{2011ApJ...730..139P,2014MNRAS.445.3055P} implements a full three-dimensional model of the BLR. The BLR emission is modelled as coming from a number of point particles that are drawn from a parametrized phase space distribution and linearly reprocess radiation from a central source. All measured parameters have a direct physical interpretation, which means that the black hole mass $M_\textrm{BH}$ and the virial factor $f$ can be determined directly. Dynamical modelling results include fully velocity-resolved transfer functions. Because the method is based on a physical model of the BLR, the response will be correlated across velocity bins. This is contrary to our implementation of regularized linear inversion, where each velocity bin is analysed individually.

Another important difference between dynamical modelling and RLI is that, because dynamical modelling traces photons through the BLR to the observer, the resulting velocity--delay maps are transfer functions representing the absolute reprocessed emission. This is contrary to RLI which subtracts the mean component of the light curves to only calculate the response in the emission-line to variations in the continuum, thus measuring response (rather than transfer) functions. However, in the current implementation of dynamical modelling by \cite{2011ApJ...730..139P}, the responsivity of the BLR gas is assumed to be constant throughout the BLR. In other words, the response of the BLR is correlated one-to-one with the emissivity distribution, regardless of the level of ionizing continuum. Because of this, the transfer functions and response functions will be directly proportional, and we can compare response maps from RLI to the velocity-resolved transfer functions of dynamical modelling directly.

% Maximum entropy
Maximum entropy methods \citep{1984MNRAS.211..111S,1991ApJ...371..541K,1991ApJ...367L...5H,1994ASPC...69...23H} work by selecting solutions to the convolution problem (Equation \ref{eq:continous_transferequation}) that simultaneously provide a good fit to the data while being as simple (smooth) as possible. The transfer functions are proposed based on a number of assumptions about the form of the solutions. This means that, like the dynamical modelling method, the derived response values will be dependent across velocity bins and the selection of the solution with the maximum entropy tends to result in very smooth velocity--delay maps. This is similar to RLI except that we impose no correlations between velocity bins, so any correlations between bins will be driven purely by the data. We compare our results to the response map for \hbeta in Arp 151 from maximum entropy \citep{2010ApJ...720L..46B} and dynamical modelling \citep{2014MNRAS.445.3073P} in Fig. \ref{fig:result-comparison-2d}. The comparisons are described in Section \ref{sec:results-arp151}.

\subsection{Results for each AGN} \label{sec:results-for-each-agn}

\begin{table*}\centering
\begin{tabular}{lcccccc}\toprule
Object   & Continuum & Emission  & Wavelength  & \multicolumn{3}{|c|}{Time delay}                                \\
         & Band      & Line      & Range (\AA) & \multicolumn{3}{|c|}{(days)}                                    \\
         &           &           &             & $\tccf$ &           $\tjav$ &             $\trli$               \\
\midrule
Arp 151  & $B$         & \halpha & 6575 - 6825 & $7.8^{+1.0}_{-1.0}$ &                     & $6.8^{+0.9}_{-1.4}$ \\
Arp 151  & $B$         & \hbeta  & 4792 - 4934 & $4.0^{+0.5}_{-0.7}$ & $3.6^{+0.7}_{-0.2}$ & $4.0^{+0.7}_{-0.8}$ \\
Arp 151  & $B$         & \hgamma & 4310 - 4393 & $1.4^{+0.8}_{-0.7}$ &                     & $3.0^{+0.8}_{-0.8}$ \\
Mrk 1310 & $B$         & \hbeta  & 4815 - 4914 & $3.7^{+0.6}_{-0.6}$ & $4.2^{+0.9}_{-0.1}$ & $2.7^{+0.3}_{-0.3}$ \\
NGC 5548 & $V$         & \hbeta  & 4706 - 5041 & $4.2^{+0.9}_{-1.3}$ & $5.5^{+0.6}_{-0.7}$ & $4.7^{+1.8}_{-1.8}$ \\
NGC 6814 & $V$         & \hbeta  & 4776 - 4936 & $6.5^{+0.9}_{-1.0}$ & $7.4^{+0.1}_{-0.1}$ & $7.3^{+1.2}_{-1.0}$ \\
SBS 1116+583A & $B$    & \hbeta  & 4797 - 4926 & $2.3^{+0.6}_{-0.5}$ & $2.4^{+0.9}_{-0.9}$ & $2.0^{+1.1}_{-0.6}$ \\
\bottomrule
\end{tabular}
\caption[The caption]{Time delays from cross-correlation ($\tccf$) are reproduced from \cite{2010ApJ...716..993B}. Time delays from \javelin ($\tjav$) are reproduced from \cite{2013ApJ...773...90G}. All time delays are given in the rest frame.}
\label{table}
\end{table*}

\subsubsection{Arp 151} \label{sec:results-arp151}

\begin{figure*}
    \includegraphics[totalheight=230pt]{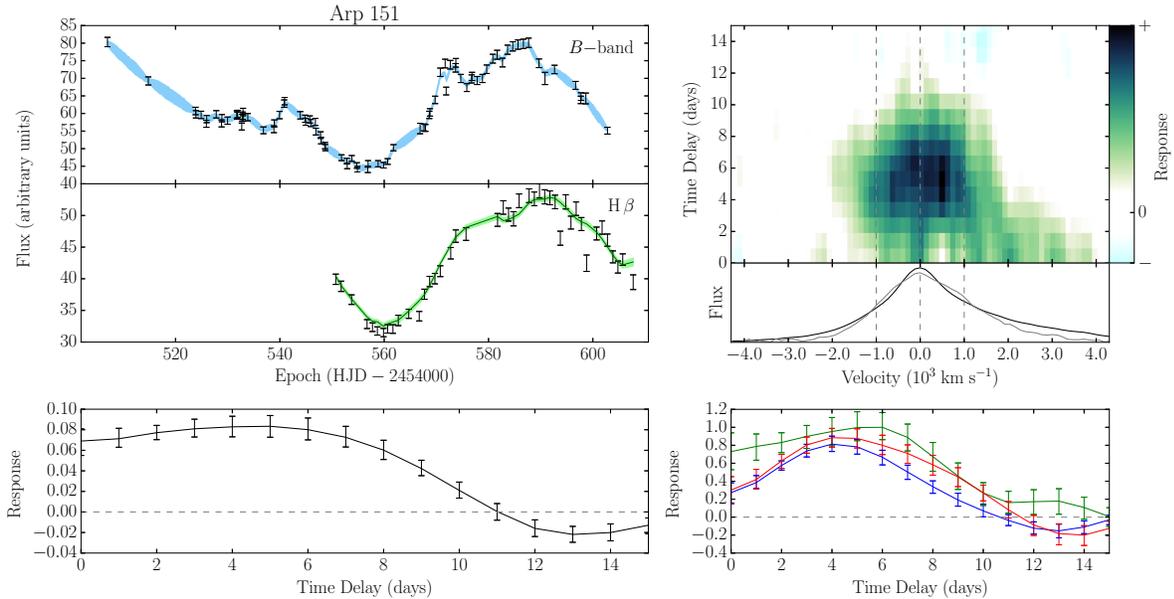}
    \caption{Results from regularized linear inversion of light curves from LAMP 2008 of the Arp 151 \hbeta emission line. The left panel shows the continuum (upper left panel, black points) and integrated emission-line light curves (middle left panel, black points) along with the integrated response function (lower left panel, black points). The blue shaded region on top of the continuum light curve shows the median and 1$\sigma$ percentiles from the Gaussian process realizations used to model the uncertainty in the data. The green line on top of the emission line light curve shows the best fitting result from RLI, with the green shaded band indicating the 1$\sigma$ percentiles. The horizontal grey dashed line in the lower left panel indicates the location of zero response. The right panel shows the velocity--delay map (upper right panel) calculated by regularized linear inversion of light curves for each wavelength bin individually. Below the velocity--delay map, the mean (black line) and RMS (grey line) spectrum is shown as a function of velocity with respect to the \hbeta emission line centre (middle-right panel). Normalized response functions for three velocity bins are shown below the velocity--delay map (lower right panel) with the bins indicated by dashed vertical lines in the velocity--delay map. The colours (blue, green, red) correspond to the Doppler shift with respect to the observer, with blue being negative velocity, green is the central velocity, and red is positive velocity. All time delays in the results figures are in the observed frame.}
    \label{fig:result-arp151}
\end{figure*}

\begin{figure*}
    \includegraphics[totalheight=230pt]{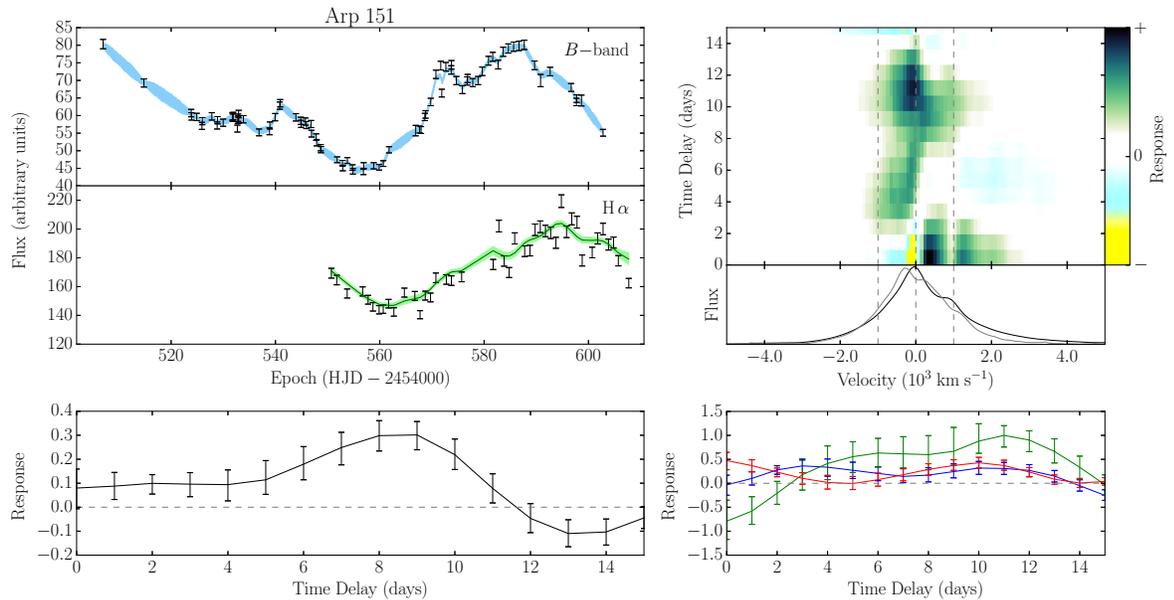}
    \caption{Same as Fig. \ref{fig:result-arp151} but for Arp 151, \halpha.}
    \label{fig:result-arp151-halpha}
\end{figure*}

\begin{figure*}
    \includegraphics[totalheight=230pt]{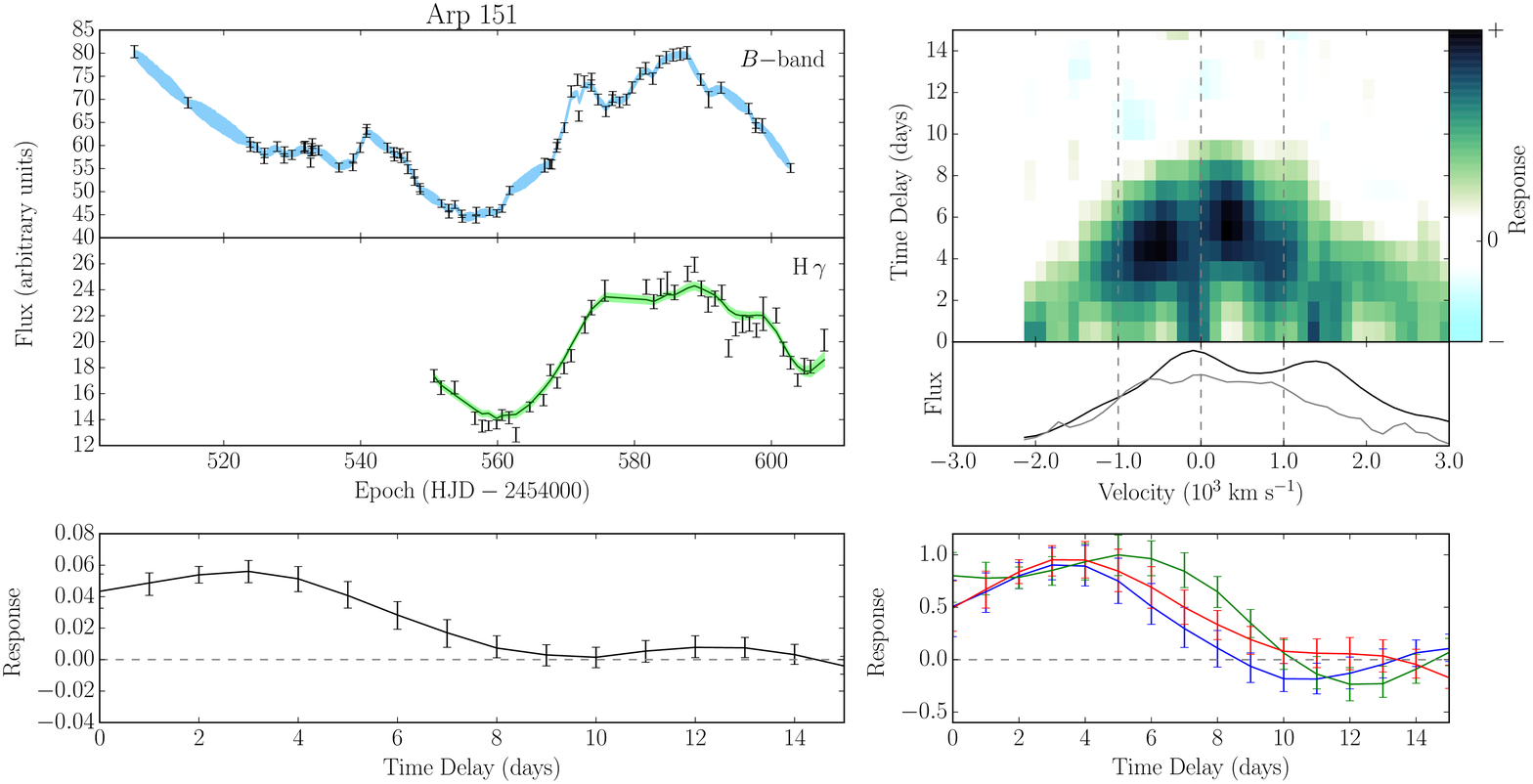}
    \caption{Same as Fig. \ref{fig:result-arp151} but for Arp 151, \hgamma.}
    \label{fig:result-arp151-hgamma}
\end{figure*}

% Light curves
Results for Arp 151 are presented in Fig. \ref{fig:result-arp151} for \hbeta and Fig. \ref{fig:result-arp151-halpha} and \ref{fig:result-arp151-hgamma} for \halpha and \hgamma respectively. Light curves for Arp 151 are the most variable and highest quality of the LAMP 2008 dataset, which is why we include only \halpha and \hgamma for this object. RLI provides decent fits to the Arp 151 light curves, with some notable outliers at later epochs (e.g. \hbeta, Fig. \ref{fig:result-arp151}). Because of the smoothing constraint imposed on the solution we do not expect our code to fit these points. This is similar to what is found in other analyses of the same data \citep{2010ApJ...720L..46B,2014MNRAS.445.3073P}. Such strong variability on short time-scales is in any case not consistent with models where the BLR has only extended emission (see discussion in Section \ref{sec:discussion-linear-response}).

% Integrated response function
The integrated response function for \hbeta (Fig. \ref{fig:result-arp151}) has a plateau from 0 days out to about 7 days where it starts to decrease. The shape of the response function is broader than what is found by maximum entropy methods \citep{2010ApJ...720L..46B}. We find a median \hbeta time delay of $\trli = 4.0^{+0.7}_{-0.8}$\,days, consistent with results from cross-correlation ($\tccf = 4.0^{+0.5}_{-0.7}$\,days) and \javelin ($\tjav = 3.6^{+0.7}_{-0.2}$\,days). The integrated response function for \halpha (Fig. \ref{fig:result-arp151-halpha}) has a flat low response out to about 6 days after which is rises to a peak response at a time delay of around 8 days, and then drops to slightly negative response after 10 days. The time delay for \halpha ($\trli = 6.8^{+0.9}_{-1.4}$) is consistent with that from cross correlation ($\tccf = 7.8^{+1.0}_{-1.0}$). The integrated response function for \hgamma (Fig. \ref{fig:result-arp151-hgamma}) shows significant response at zero delay with a slight rise to a peak at 3 days after which the response drops to zero beyond 8 days. The time delay from RLI ($\trli = 3.0^{+0.8}_{-0.8}$) is somewhat larger, but consistent with, the result from cross correlation ($\tccf = 1.4^{+0.8}_{-0.7}$).

% Velocity-resolved response maps
The velocity--delay map for \hbeta (Fig. \ref{fig:result-arp151}) shows that the bulk of the response is centred on the emission line with a time delay of about 5 days, similar to the median delay calculated from the integrated response. There seems to be an area of increased response redward of the line centre from $0 - 1000\,\textrm{km}/\textrm{s}$. The velocity--delay map for \halpha (Fig. \ref{fig:result-arp151-halpha}) shows a strong response around 10 days centred on the emission line and extending to lower time delays. The origin of the prompt response redward of the line centre is unclear. It may be a spurious feature due to numerical effects, a poorly subtracted continuum (the continuum for \halpha was subtracted using a linear fit rather than spectral decomposition as for \hbeta, see Section \ref{sec:emission-line-lightcurves}), or perhaps due to residual contamination from \nii. We do not find evidence for a blueward plume from 15 to 20 days (not shown) as seen in the maximum entropy maps in \cite{2010ApJ...720L..46B}. The velocity--delay map for \hgamma (Fig. \ref{fig:result-arp151-hgamma}) shows a broad response with the longest delays at the centre of the line, and progressively decreasing time delays in the wings. We find additional prompt response at the centre of \hgamma.

The decrease of the median time delay in Arp 151 from \halpha to \hbeta and \hgamma has been previously observed (\citealt{2010ApJ...720L..46B}, see \citealt{2009NewAR..53..140G} for a review), and may be an effect of the varying optical depth for the Balmer lines. Because the optical depth is larger for the transitions between lower excitation states, \halpha photons will have a harder time escaping the BLR without being absorbed. This results in the \halpha emission being predominantly directed back towards the ionizing source at the centre, whereas the lines formed by the higher excitation states, \hbeta and \hgamma, are progressively more isotropic. That is one possible mechanism by which the median time delay can decline in higher excitation emission lines, while all the Balmer lines originate from hydrogen gas located at similar distances from the ionizing source.

% Comparison - Misty bins
Fig. \ref{fig:result-misty-bins} shows a comparison between RLI and velocity-resolved cross-correlation time delays by \cite{2009ApJ...705..199B} for a number of velocity bins across the \hbeta emission line. We find excellent agreement with cross-correlation for all velocity bins and confirm the signature of prompt response redward of the \hbeta line centre, while the bulk of the response is at the centre of the emission line at a time delay of around 5 days.

% Comparison - Dynamical modelling and maximum entropy
Fig. \ref{fig:result-comparison-2d} shows velocity--delay maps for \hbeta from dynamical modelling, maximum entropy, and RLI. All methods find prompt response in the red wing of the \hbeta emission line. This is a characteristic feature of BLR models with free-falling gas or a disk of gas containing a hot spot \citep[see][]{1991ApJ...379..586W,2010ApJ...720L..46B}. Some models also produce prompt red-side response for outflowing gas at particular observed orientations \citep{1997ApJ...479..200B}. The result from RLI has a stronger response at the line centre, and a weaker prompt response in the red wing compared to dynamical modelling and maximum entropy. In addition RLI finds prompt response at the line centre, which is not seen in the maximum entropy maps.

\subsubsection{Mrk 1310}

\begin{figure*}
    \includegraphics[totalheight=230pt]{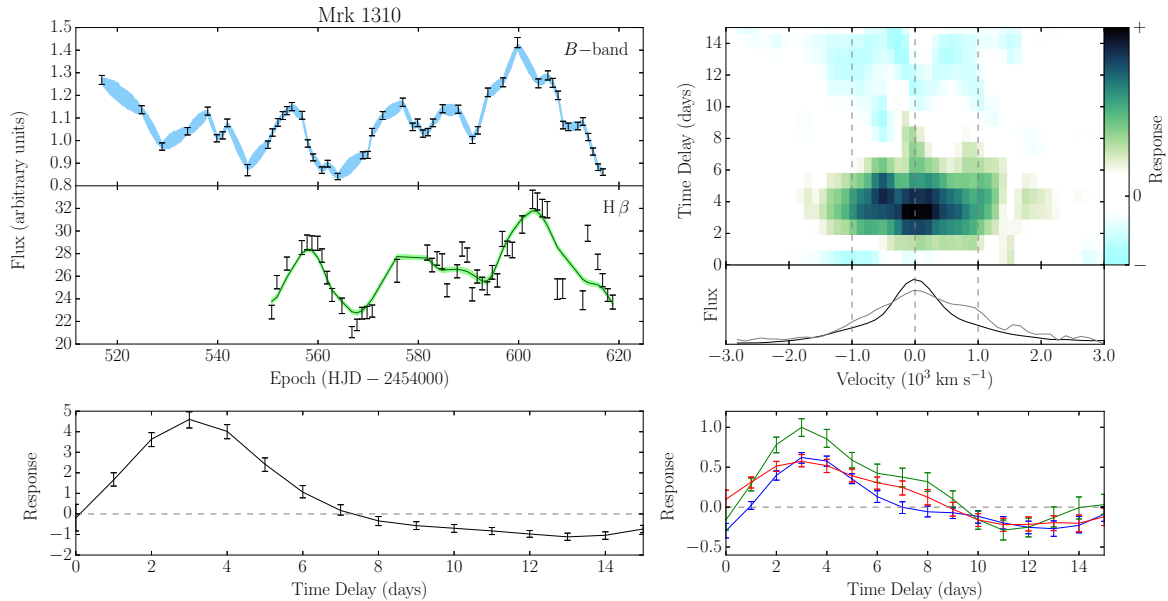}
    \caption{Same as Fig. \ref{fig:result-arp151} but for Mrk 1310, \hbeta.}
    \label{fig:result-mrk1310}
\end{figure*}

% Light curves
Results for Mrk 1310 are presented in Fig. \ref{fig:result-mrk1310}. There is more scatter in the light curves of Mrk 1310 compared to Arp 151. Because of the smoothing constraint, RLI is not able to fit the emission line light curve exactly. This is to be expected, as the method mainly picks up large scale variability to avoid problems with noise on shorter time-scales. The scatter in the \hbeta light curve towards the end of the campaign is in any case difficult to reconcile with a simple linear response to the continuum light curve variations near the same epochs.

% Integrated response function
The integrated response function for \hbeta is more peaked than that of Arp 151. The scatter in the derived response functions from our error analysis is also smaller, indicating that the inversion is stable under perturbations in the input light curves. The median time delay of the response function is $\trli = 2.7^{+0.3}_{-0.3}$\,days. This delay is slightly shorter than, but consistent with, the cross-correlation result ($\tccf = 3.7^{+0.6}_{-0.6}$\,days), while the result from \javelin ($\tjav = 4.2^{+0.9}_{-0.1}$\,days) is longer by 1 day compared to the time delay from RLI. The integrated response function dips slightly below zero at longer time delays, which could indicate a negative response in the emission-line, although we do not find this to be significant (see discussion in Section \ref{sec:negative-response}).

% Velocity-resolved response maps
% Comparison - Misty bins
% Comparison - Dynamical modelling and maximum entropy
The velocity-resolved response map shows a strong response at the line centre with a time delay around 3 days, similar to the integrated time delay. The width of the response is about $\pm 1000$\,\kms, similar to Arp 151. The delay is fairly constant across the \hbeta line. This result agrees well with velocity-resolved cross-correlation that shows a nearly flat response as a function of velocity, with slightly shorter delays in the wings at $\pm 1000$\,\kms (Fig. \ref{fig:result-misty-bins}). It is also consistent with dynamical modelling results, with a peak response centred on the emission line at fairly short time delays (Fig. \ref{fig:result-comparison-2d}, middle left panel).

\subsubsection{NGC 5548}

\begin{figure*}
    \includegraphics[totalheight=230pt]{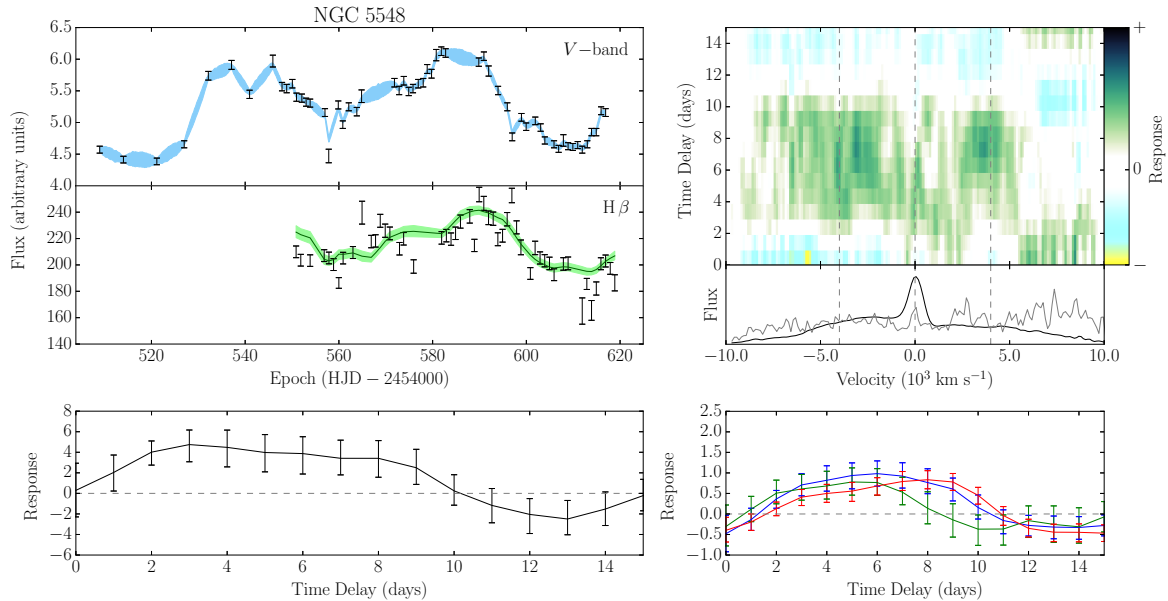}
    \caption{Same as Fig. \ref{fig:result-arp151} but for NGC 5548, \hbeta.}
    \label{fig:result-ngc5548}
\end{figure*}

% Light curves
Results for NGC 5548 are presented in Fig. \ref{fig:result-ngc5548}. The \hbeta light curve of NGC 5548 has more scatter relative to variability amplitude compared to Arp 151 and Mrk 1310. Consequently, RLI has a more difficult time fitting this object.

% Integrated response function
The integrated response function increases from zero delay up to around 3 days and then has a slowly decreasing plateau out to about 8 days after which it drops off. This yields a median time delay of $\trli = 4.7^{+1.8}_{-1.8}$\,days, consistent within 1$\sigma$ with cross-correlation ($\tccf = 4.2^{+0.9}_{-1.3}$\,days) and \javelin ($\tjav = 5.5^{+0.6}_{-0.7}$\,days).

% Velocity-resolved response maps
% Comparison - Misty bins
% Comparison - Dynamical modelling and maximum entropy
The \hbeta emission line in NGC 5548 is significantly broader compared to the other objects analysed and the variability across the line shows more irregular structure, as is seen in the velocity-resolved response map in the upper right panel of Fig. \ref{fig:result-ngc5548}. We find a somewhat weak response at the line centre, which peaks at around 5 days. On either side of the line centre at $\pm 5000$\,\kms we find stronger isolated response features with peak time delays around 8 days. The velocity-resolved time delays agree well with cross-correlation (Fig. \ref{fig:result-misty-bins}). Dynamical modelling and RLI both show a level of prompt response in the red wing of \hbeta, while the rest of the maps show little similarity (Fig. \ref{fig:result-comparison-2d}).

\subsubsection{NGC 6814}

\begin{figure*}
    \includegraphics[totalheight=230pt]{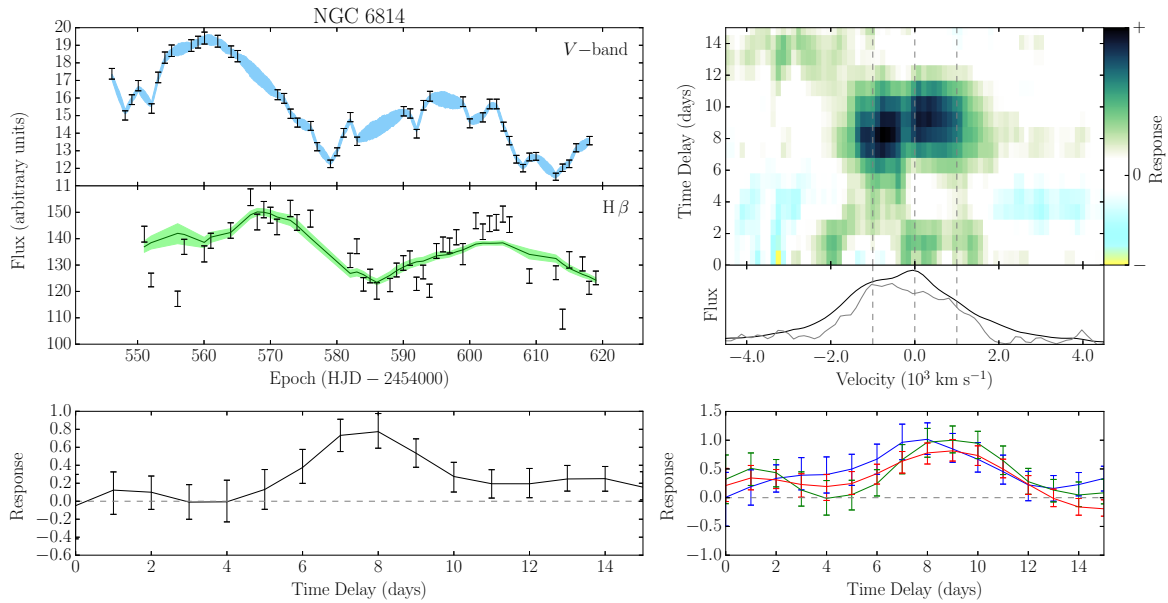}
    \caption{Same as Fig. \ref{fig:result-arp151} but for NGC 6814, \hbeta.}
    \label{fig:result-ngc6814}
\end{figure*}

% Light curves
Results for NGC 6814 are presented in Fig. \ref{fig:result-ngc6814}. The light curves of NGC 6814 show an appreciable amount of variability. The continuum $V$-band light curve has a clear broad peak at $\textrm{HJD} - 2454000 = 560\,\textrm{days}$ after which it drops off to a slowly rising plateau that extends to shortly after $\textrm{HJD} - 2454000 = 600\,\textrm{days}$. The emission line light curve has the same over-all trend as the continuum, but instead of having a plateau after the initial peak, it rises to almost the same level as the initial peak. Because of this, RLI has a difficult time matching the first and second peaks in the emission line light curve, explaining why the fit is underestimated for the second peak.

% Integrated response function
The integrated response function peaks around 8 days and has broad flat wings to either side (Fig. \ref{fig:result-ngc6814}, lower left panel). The median time delay of $\trli = 7.3^{+1.2}_{-1.0}$\,days is consistent with results from cross-correlation ($\tccf = 6.5^{+0.9}_{-1.0}$\,days) and \javelin ($\tjav = 7.4^{+0.1}_{-0.1}$\,days).

% Velocity-resolved response maps
% Comparison - Misty bins
% Comparison - Dynamical modelling and maximum entropy
The velocity-resolved response map for NGC 6814 shows a clear isolated response peak centred on the \hbeta emission line (Fig. \ref{fig:result-ngc6814}, upper right panel). There is a somewhat peculiar dip in the response map just blueward ($\sim -100$\,\kms) of the line centre. Further away at around $\pm 500$\,\kms the response peaks on either side of the line centre and then drops off to zero beyond $\pm 1500$\,\kms. The velocity-resolved time delays agree well with cross-correlation at centre of the \hbeta line (Fig. \ref{fig:result-misty-bins}). At lower velocities we find a slightly longer time delay compared to the cross-correlation result. At higher velocities, blueward of the line, RLI was unable to recover a time delay due to a lack of emission line response. Looking at the fully resolved response map from RLI in Fig. \ref{fig:result-ngc6814} we see that RLI calculates little response in the wings beyond $\pm 1500$\,\kms, which might explain the discrepancy with the cross-correlation results. Dynamical modelling of NGC 6814 tends to prefer shorter time delays than what is found using other methods (see Fig. \ref{fig:result-comparison-2d}). The highest velocity bin failed to produce a time delay estimate due to a noisy response function.

\subsubsection{SBS 1116+583A}

\begin{figure*}
    \includegraphics[totalheight=230pt]{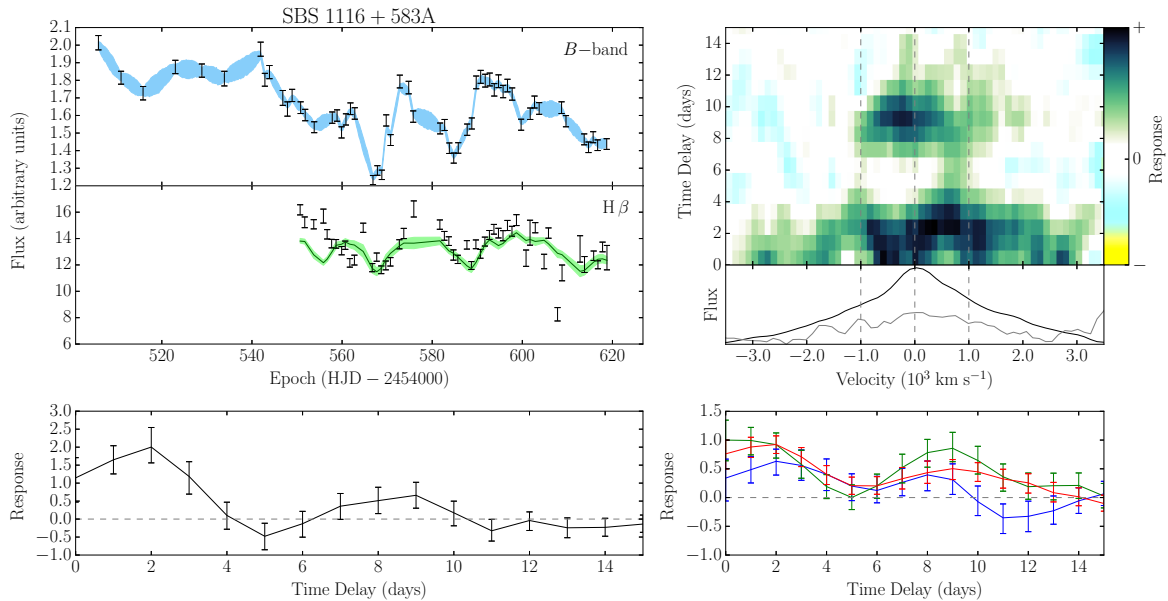}
    \caption{Same as Fig. \ref{fig:result-arp151} but for SBS 1116+583A, \hbeta.}
    \label{fig:result-sbs1116}
\end{figure*}

% Light curves
Results for SBS 1116+583A are presented in Fig. \ref{fig:result-sbs1116}. The emission line light curve of SBS 1116+583A lacks significant features above the noise level. As a result the RLI fit does not match the emission line light curve well. The results for this object, including the velocity--delay map, should thus be interpreted with caution.

% Integrated response function
The integrated response function prefers a relatively short time delay of $\trli = 2.0^{+1.1}_{-0.6}$\,days, which is slightly shorter, but fully consistent with cross-correlation ($\tccf = 2.3^{+0.6}_{-0.5}$\,days) and \javelin ($\tjav = 2.4^{+0.9}_{-0.9}$\,days).

% Velocity-resolved response maps
% Comparison - Misty bins
% Comparison - Dynamical modelling and maximum entropy
The velocity-resolved response map for SBS 1116+583A (Fig. \ref{fig:result-sbs1116}) shows a multimodal response map with a broad response component at short time delays, around 2 days, and an additional isolated response at 10 days at the centre of the \hbeta line. This is a good example of the flexibility of RLI to reconstruct complicated velocity--delay maps. The velocity--delay map shows some resemblance to models in which the BLR is confined to an isotropically illuminated disk \citep[see Fig. 5 in][]{2010ApJ...720L..46B}.

\begin{figure*}
    \includegraphics[width=1.0\linewidth]{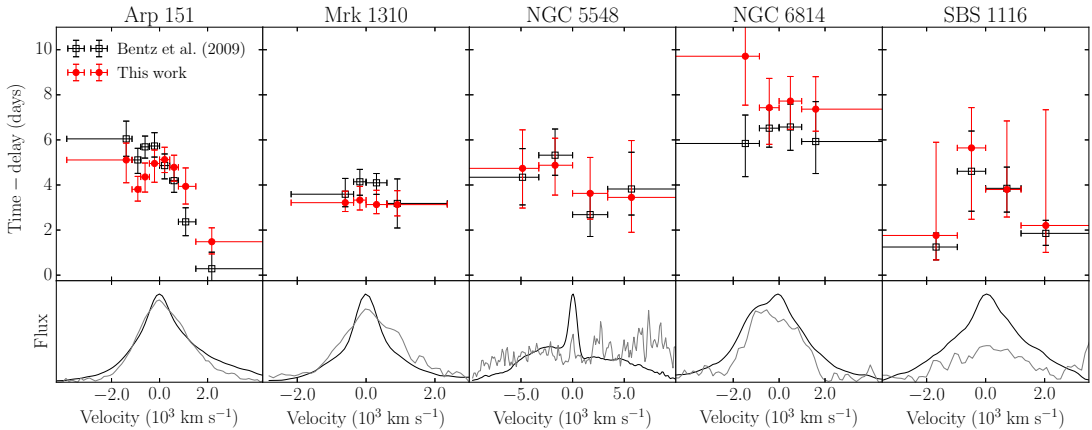}
    \caption{Comparison of \hbeta time delays from regularized linear inversion with cross-correlation results from \protect\cite{2009ApJ...705..199B}. The upper panels show time delays as a function of velocity with respect to the \hbeta emission line centre. Red filled circles with error bars show time delays from this paper. Black open squares with error bars are from \protect\cite{2009ApJ...705..199B}. Light curves are obtained by integrating the emission in each velocity bin indicated by the horizontal error-bars. The time delays from RLI are calculated as the median of the positive part of the response (see Section \ref{sec:the-time-delay}). The lower panels show the mean (black) and RMS (grey) spectra. All time delays in this figure are in the observed frame.}
    \label{fig:result-misty-bins}
\end{figure*}

It is not very meaningful to calculate a single time delay from a multimodal response function, even so we apply the same procedure for determining the time delay as for the previous objects (see Section \ref{sec:the-time-delay}). The resulting velocity-resolved time delays agree well with cross-correlation, showing longer time delays at the centre of the emission line and shorter time delays in the wings (Fig. \ref{fig:result-misty-bins}). We note that the time delay calculated close to the \hbeta line centre fall between the two modes in the response map, and is thus not well constrained. We include these time delays for comparison with the velocity-resolved cross-correlation, but they should be interpreted with caution, as suggested by the large error-bars. The lowest velocity bin failed to produce a time delay estimate due to a noisy response function.

\begin{figure*}
    \includegraphics[width=1.0\linewidth]{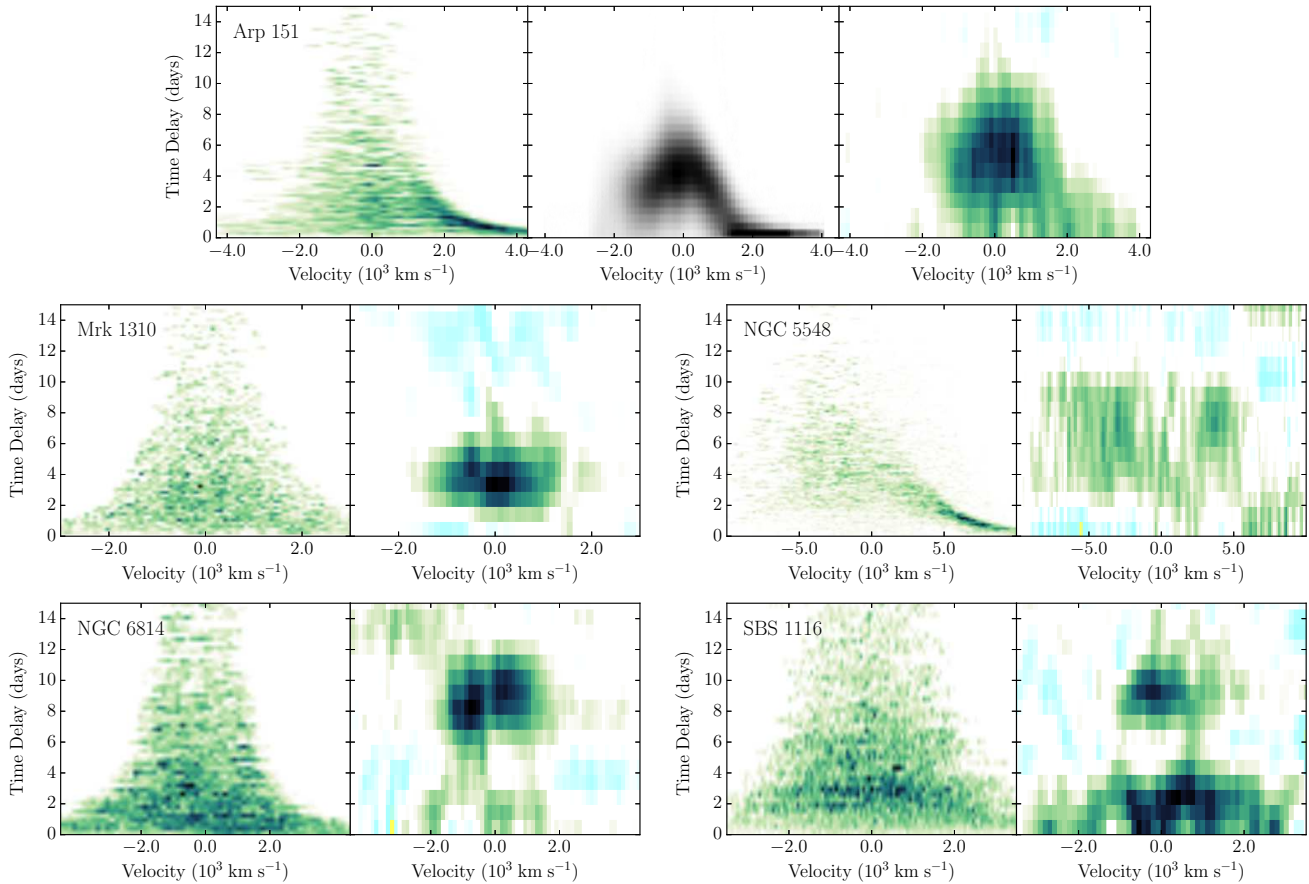}
    \caption[]{Comparison of velocity--delay maps from regularized linear inversion to maximum likelihood for Arp 151 \citep{2010ApJ...720L..46B} and dynamical modelling for all objects \citep{2014MNRAS.445.3073P}. All three methods used the same dataset from LAMP 2008 (see Section \ref{sec:data}). Regularized linear inversion and dynamical modelling used spectral decomposition for continuum subtraction, while the maximum likelihood analysis used a linear fit to remove the continuum level below the line.}
    \label{fig:result-comparison-2d}
\end{figure*}

The results for SBS 1116+583A from RLI are in general agreement with dynamical modelling, although dynamical modelling does not reproduce the same strong response at $\tau = 10$\,days and generally prefers shorter time delays at the line centre (see Fig. \ref{fig:result-comparison-2d}).

% ============================================================
% ============================================================
% ============================================================
% DISCUSSION
\section{Discussion} \label{sec:discussion}
% ============================================================

\subsection{Assumption of a constant linear response} \label{sec:discussion-linear-response}
Long temporal baseline multi-epoch observations of AGN reveal significant changes in the broad emission line equivalent widths \citep[e.g.][]{1990ApJ...357..338K,1992AJ....103.1084P,2003ApJ...587..123G,2004MNRAS.352..277G} and line profiles \citep[e.g.][]{1996ApJ...466..174W,2001ApJ...554..245S,2007ApJ...668..708S} on time-scales of months to years. These effects could possibly account for some of the discrepancies found when matching longer time-scale variability patterns using linear reverberation mapping, such as the inability of RLI to match the second peak in the emission line light curve in NGC 6814 (Fig. \ref{fig:result-ngc6814}).

In addition to long time-scale non-linearities, some light curves of the LAMP 2008 dataset show high-variability features (e.g. late epochs in Mrk 1310, Fig. \ref{fig:result-mrk1310}). If these features are associated with the BLR, they are inconsistent with models in which the BLR emission comes from an extended region. This remains true even after exclusion of potentially unreliable spectra in the LAMP 2008 dataset, as described in section \ref{sec:emission-line-lightcurves}. Whether these outliers are due to systematic measurement uncertainties affecting individual epochs, an indication of unknown processes in the BLR, or a combination of these, is unclear.

Whatever the origin of non-linear features in the emission line light curves, current reverberation mapping techniques, such as cross-correlation, maximum entropy, and RLI, will be insufficient to describe them. Even so these methods remain valuable tools for testing models of the BLR and investigate departures from linearity in AGN light curves. Implementation of photoionization physics into dynamical modelling codes may be a way to probe non-linear processes in the BLR.

\subsection{Ionizing versus observed continuum}
Studies suggest a non-negligible time delay between the ionizing UV continuum and the optical continuum on the order of 1 day in NGC 7469 and NGC 5548 \citep[e.g.][]{1998ApJ...500..162C,2014MNRAS.444.1469M,2015ApJ...806..129E}. This means that the $V$ and $B$ band continuum light curves originate from a region comparable in size to some of the shorter time delays found for Balmer emission lines. This extended continuum emission region will introduce geometric smoothing in the optical continuum light curves used for reverberation mapping. If the delay between the ionizing continuum and the measured optical continuum is constant on time-scales of reverberation mapping campaigns, linear reverberation methods can still be applied, but the interpretation of the response functions will have to be revised to include not just the geometry of the BLR, but also the geometry of the extended optical continuum emission region. It will be interesting to extend this analysis in the future by performing RLI of optical emission line variability as driven by UV continuum to assess the magnitude of these systematic effects.

\subsection{Response function errors}
% Smoothing and correlated errors
The smoothing condition used to regularize the inversion problem introduces correlations in the response functions by linking nearby points through the first-order derivative. This correlation is an implicit assumption of the method and it reflects our belief that the emission line light curve is produced by a superposition of photons emitted from a fairly homogeneous distribution of gas in the BLR. The error bars on the response functions should thus not be interpreted as errors on the individual points of the response function, but rather as an indication of the range of solutions we get by re-interpolating the continuum light curve and re-sampling the emission line light curve.

% Smoothing in the velocity-domain
In our current implementation of RLI no correlation in response is imposed between response velocity bins. This can be witnessed by the vertical streaks in the presented response maps. Not imposing a correlation has the advantage that any correlation between velocity-bins in the final response maps must be driven by the data, or at least systematic effects correlated with the emission lines. It seems natural to assume that the BLR emissivity is correlated in position as well as velocity space, motivating a smoothing constraint in the velocity domain, as is also assumed in dynamical modelling and maximum entropy methods. Because we found the velocity-bins to be naturally correlated in the response maps, and in order to keep the method fast and simple, we decided not to impose any smoothing in the velocity-domain.

\subsection{Negative response values} \label{sec:negative-response}
Some models suggest that negative values occur naturally in AGN response functions \citep{1993ApJ...404..570S,1993MNRAS.263..149G,2014MNRAS.444...43G}, and negative response values have recently been observed in X-rays \citep{2007MNRAS.382..985M,2009Natur.459..540F}. Negative response can arise in different ways, such as if an increase in ionizing flux causes part of the responding gas to become fully ionized, or if the BLR structure changes on a dynamical time-scale, such as if the continuum decreases while new material falls into the BLR. Note that in both of these examples the BLR structure changes on time-scales of the same order as the time delay. This means that the assumption of a constant linear response breaks down and more general methods, such as dynamical modelling, are required to account for the time varying part of the response.

An advantage of RLI over other reverberation mapping methods, is that it does not require the response function to be everywhere positive. We find some preference for negative response in some of the objects analysed (e.g. Mrk 1310, Fig. \ref{fig:result-mrk1310}). While none of the response functions show strong evidence for negative response these results may be an indication that more complicated processes are taking place in the BLR, than what can be contained in a simple constant linear response model. However, negative values occur naturally as artefacts when linear inversion methods are applied to discrete and noisy data, even if the true response is everywhere positive. This is somewhat in analogy to aliasing distortions when dealing with discrete Fourier transforms. These artefacts are seen as ringing effects, such as observed in the inversion of the simulated 1D data (see Fig. \ref{fig:sim-1d}). It is not possible to impose positivity while keeping an analytical expression for the response function \citep{1994PASP..106.1091V}, and since simplicity and flexibility were the main motivations for using RLI, we leave the positivity constraint to other methods such as maximum entropy and dynamical modelling. With these considerations in mind we emphasize that all the negative response values found in our analysis of the LAMP 2008 data are of low statistical significance. In particular, the integrals of the response functions are positive for all objects analysed.

% ============================================================
% ============================================================
% ============================================================
% CONCLUSION
\section{Summary and conclusion} \label{sec:summary-and-conclusion}

Building on previous work by \cite{1995ApJ...440..166K}, we develop a new method for reverberation mapping based on regularized linear inversion that includes statistical modelling of the AGN continuum light curves. In this implementation, response function errors are evaluated using a combination of Gaussian process fitting of the continuum light curves as well as bootstrap resampling of the emission line light curves. Regularized linear inversion has the advantage over other reverberation mapping methods that it makes few assumptions about the shape of the response function, and it allows for negative values in the response function. In addition, because the method is based on an analytical solution to the transfer equation, it is very computationally efficient. The scale of regularization is a function of the data only, such that no user-input is required, except for the time baseline of the analysis. This means that flexible response functions can be calculated quickly and consistently.

We test the method on simulated data and show that it is able to recover unimodal and multimodal response functions as well as response functions with negative values (see Fig. \ref{fig:sim-1d}). We further test the method on data simulated using dynamical modelling and show that it is able to recover 2-dimensional velocity--delay maps (see Fig. \ref{fig:sim-2d}).

We apply RLI to light curves of five nearby Seyfert 1 galaxies, taken by the LAMP 2008 collaboration, and present time delays, integrated response functions and velocity--delay maps for the H$\beta$ line in all objects, as well as H$\alpha$ and H$\gamma$ for Arp 151. This is the first time a reverberation mapping method allowing for negative response has been applied to a large dataset. Our results are in good agreement with previous studies based on cross-correlation and dynamical modelling, offering a powerful corroboration of the assumptions underlying these reverberation mapping methods.

The main physical results of this work can be summarized as follows:

\begin{enumerate}
    \item Despite using a very different method for calculating time delays we find that our results are in good agreement with results from cross correlation.
    \item We find asymmetric response of the H$\beta$ emission line in Arp 151, with prompt response in the red wing of the emission line, consistent with models that include bulk gas flows in the BLR.
    \item We confirm previous studies finding that lines originating from high excitation states, such as \hgamma, have shorter time delays compared to lines originating from lower excitation states, such as \halpha and \hbeta.
    \item While some objects, such as Arp 151 and Mrk 1310, show a degree of negative response at longer time delays, we find no conclusive evidence for negative response values in any of the objects.
\end{enumerate}

This work demonstrates that regularized linear inversion is a valuable tool for reverberation mapping and a worthwhile complementary method for analysing high quality reverberation mapping datasets. The recent multi-wavelength reverberation mapping campaign of NGC 5548 \citep{2015ApJ...806..128D,2015ApJ...806..129E}, including simultaneous ground-based monitoring in the optical and spectroscopy in the ultraviolet with the Cosmic Origins Spectrograph on the \emph{Hubble Space Telescope}, will provide exceptionally high quality data and thereby a great opportunity to test the RLI method and its underlying assumptions.

% ============================================================
% ============================================================
% ============================================================
% ACKNOWLEDGMENTS
\section*{acknowledgements}
% ============================================================

We thank the referee, Rick Edelson, for comments and suggestions that led to substantial improvements of the manuscript.
We are grateful to Julian Krolik, Mike Goad, Kirk Korista and Brendon Brewer for enlightening input, as well as Anthea King and Darach Watson for discussions leading to the improvement of this work.
The Dark Cosmology Centre (DARK) is funded by the Danish National Research Foundation.
AP acknowledges support from the NSF through the Graduate Research Fellowship Program and from the University of California Santa Barbara through a Dean's Fellowship.
TT acknowledges support from the Packard foundation in the form of a Packard Fellowship.
TT thanks the American Academy in Rome and the Observatory of Monteporzio Catone for their kind hospitality during the writing of this manuscript.
Research by TT is supported by NSF grant AST-1412693 (New Frontiers in Reverberation Mapping).
DP acknowledges support through the EACOA Fellowship from The East Asian Core Observatories Association, which consists of the National Astronomical Observatories, Chinese Academy of Science (NAOC), the National Astronomical Observatory of Japan (NAOJ), Korean Astronomy and Space Science Institute (KASI), and Academia Sinica Institute of Astronomy and Astrophysics (ASIAA).
Research by AJB is supported by NSF grants AST-1108835 and AST-1412693.
Research by MCB is supported by NSF CAREER grant AST-1253702.
This work makes use of the Lick AGN Monitoring Project 2008 dataset supported by NSF grants AST-0548198 (UC Irvine), AST-0607485 and AST-0908886 (UC Berkeley), AST-0642621 (UC Santa Barbara), and AST-0507450 (UC Riverside).
This work makes use of the {\sc{eigen}} header library \citep{eigenweb}. All figures are produced in {\sc{matplotlib}} \citep{Hunter:2007}.

\clearpage

% ============================================================
% REFERENCES
% ============================================================
\bibliographystyle{mn2e}
\bibliography{regularized-inversion-paper1}

\end{document}